\documentclass[usegraphicx,usenatbib]{mn2e}
\usepackage{color,times,float,hyperref}

\newcommand{\Msol}{\ensuremath{M_\odot}}

\newcommand{\etal}{et al.}
\newcommand{\aj}{AJ}
\newcommand{\aanda}{A\&A}
\newcommand{\aandas}{A\&AS}
\newcommand{\apj}{ApJ}

\newcommand{\apjs}{ApJS}
\newcommand{\apss}{Ap\&SS}
\newcommand{\astroph}[1]{arXiv:astro-ph/{#1}}
\newcommand{\arxiv}[1]{arXiv:{#1}}
\newcommand{\mnras}{MNRAS}
\newcommand{\pasp}{PASP}

\newcommand{\prl}{Phys. Rev. Lett.}

\newcommand*{\rom}[1]{\romannumeral #1}
\newcommand{\ion}[2]{{#1}~\textsc{\rom{#2}}}

\newcommand{\ewna}{\ensuremath{EW(\mathrm{\ion{Na}{1}~D)}}}
\newcommand{\nickel}{{\ensuremath{^{56}\mathrm{Ni}}}}
\newcommand{\cobalt}{{\ensuremath{^{56}\mathrm{Co}}}}

\newcommand{\vSi}{\ensuremath{v_{\mathrm{Si}}}}
\newcommand{\vdot}{\ensuremath{\dot{v}_{\mathrm{Si}}}}
\newcommand{\kms}{\ensuremath{\mathrm{km~s}^{-1}}}

\newcommand{\MNi}{\ensuremath{M_{\mathrm{Ni}}}}
\newcommand{\MWD}{\ensuremath{M_\mathrm{ej}}}
\newcommand{\vKE}{\ensuremath{v_\mathrm{KE}}}
\newcommand{\aNi}{\ensuremath{a_\mathrm{Ni}}}
\newcommand{\trise}{\ensuremath{t_\mathrm{rise}}}

\newcommand{\revised}[1]{{\textcolor{black} {#1}}}
\newcommand{\nb}[1]{\ensuremath{^{#1}}}
\newcommand{\code}[1]{{\sc #1}}

\begin{document}


\title[UV + optical observations of LSQ12gdj]
      {Early ultraviolet emission in the Type Ia supernova LSQ12gdj: \\
       No evidence for ongoing shock interaction}

\author[Scalzo et al.]
{
    R.~A.~Scalzo\nb{1,2}\thanks{Email: rscalzo@mso.anu.edu.au},
    M.~Childress\nb{1,2},
    B.~Tucker\nb{1,3},
    F.~Yuan\nb{1,2},
    B.~Schmidt\nb{1,2},
    P.~J.~Brown\nb{4}, \newauthor
    C.~Contreras\nb{5},
    N.~Morrell\nb{6},
    E.~Hsiao\nb{6},
    C.~Burns\nb{7},
    M.~M.~Phillips\nb{6},
    A.~Campillay\nb{6}, \newauthor
    C.~Gonzalez\nb{6},
    K.~Krisciunas\nb{4},
    M.~Stritzinger\nb{5},
    M.~L.~Graham\nb{3,8},
    J.~Parrent\nb{8,9}, \newauthor
    S.~Valenti\nb{8,10},
    C.~Lidman\nb{11},
    B.~Schaefer\nb{12},
    N.~Scott\nb{13},
    M.~Fraser\nb{14,15},
    A.~Gal-Yam\nb{16}, \newauthor
    C.~Inserra,\nb{14},
    K.~Maguire\nb{17},
    S.~J.~Smartt\nb{14},
    J.~Sollerman\nb{18},
    M.~Sullivan\nb{19},
    F.~Taddia\nb{18}, \newauthor
    O.~Yaron\nb{16},
    D.~R.~Young\nb{14},
    S.~Taubenberger\nb{20},
    C.~Baltay\nb{21},
    N.~Ellman\nb{21},
    U.~Feindt\nb{22}, \newauthor
    E.~Hadjiyska\nb{21},
    R.~McKinnon\nb{21},
    P.~E.~Nugent\nb{3,23},
    D.~Rabinowitz\nb{21},
    and E.~S.~Walker\nb{21}
    \\
    \nb{1} Research School of Astronomy and Astrophysics,
           Australian National University,
           Canberra, ACT 2611, Australia \\
    \nb{2} ARC Centre of Excellence for All-Sky Astrophysics (CAASTRO) \\
    \nb{3} Department of Astronomy, University of California, Berkeley,
           B-20 Hearst Field Annex \#3411, Berkeley, CA 94720-3411, USA \\
    \nb{4} George P. and Cynthia Woods Mitchell Institute
           for Fundamental Physics and Astronomy,
           Department of Physics and Astronomy, Texas A\&M University, \\
           4242 TAMU, College Station, TX 77843, USA \\
    \nb{5} Department of Physics and Astronomy, Aarhus University,
           Ny Munkegade 120, DK-8000 Aarhus C, Denmark \\
    \nb{6} Carnegie Observatories, Las Campanas Observatory,
           La Serena, Chile \\
    \nb{7} Observatories of the Carnegie Institution for Science,
           Pasadena, CA 91101, USA \\
    \nb{8} Las Cumbres Observatory Global Telescope Network,
           6740 Cortona Dr., Suite 102, Goleta, CA 93117, USA \\
    \nb{9} Department of Physics and Astronomy, Dartmouth College,
            Hanover, NH 03755, USA \\
    \nb{10} Department of Physics, University of California, Santa Barbara,
            Broida Hall, Mail Code 9530, Santa Barbara, CA 93106-9530, USA \\
    \nb{11} Australian Astronomical Observatory, PO Box 296,
            Epping, NSW 1710, Australia \\
    \nb{12} Department of Physics and Astronomy, Louisiana State University,
            Baton Rouge, Louisiana 70803, USA \\
    \nb{13} Center for Astrophysics \& Supercomputing,
            Swinburne University of Technology, PO Box 218,
            Hawthorn, VIC 3122, Australia \\
    \nb{14} Astrophysics Research Centre, School of Mathematics and Physics,
            Queen's University Belfast, Belfast, BT7 1NN, UK \\
    \nb{15} Institute of Astronomy, University of Cambridge, Madingley Road,
            Cambridge, CB3 0HA, UK \\
    \nb{16} Department of Particle Physics and Astrophysics,
            The Weizmann Institute of Science, Rehovot 76100, Israel \\
    \nb{17} European Southern Observatory, Karl-Schwarzschild-Str. 2,
            85748 Garching bei M\"unchen, Germany \\
    \nb{18} The Oskar Klein Centre, Department of Astronomy,
            Stockholm University, AlbaNova, 10691 Stockholm, Sweden \\
    \nb{19} School of Physics and Astronomy, University of Southampton,
            Southampton, SO17 1BJ, UK \\
    \nb{20} Max-Planck-Institut f\"ur Astrophysik, Karl-Schwarzschild-Str. 1,
            85741 Garching bei M\"unchen, Germany \\
    \nb{21} Department of Physics, Yale University,
           New Haven, CT 06520-8121, USA \\
    \nb{22} Physikalisches Institut, Universität Bonn, Nussallee 12, 53115
            Bonn, Germany \\
    \nb{23} Computational Cosmology Center, Computational Research Division,
            Lawrence Berkeley National Laboratory,
            1 Cyclotron Road MS 50B-4206, \\ Berkeley, CA 94720, USA \\
}

\maketitle

\vspace{-1in}

\begin{abstract}
We present photospheric-phase observations of LSQ12gdj, a slowly-declining,
UV-bright Type~Ia supernova.  Classified well before maximum light, LSQ12gdj
has extinction-corrected absolute magnitude $M_B = -19.8$, and pre-maximum
spectroscopic evolution similar to SN~1991T and the super-Chandrasekhar-mass
SN~2007if.  We use ultraviolet photometry from \emph{Swift}, ground-based
optical photometry, and corrections from a near-infrared photometric template
to construct the bolometric (1600--23800~\AA) light curve out to 45~days
past $B$-band maximum light.  We estimate that LSQ12gdj produced
$0.96 \pm 0.07$~\Msol\ of \nickel, with an ejected mass near or slightly
above the Chandrasekhar mass.  As much as 27\% of the flux at the earliest
observed phases, and 17\% at maximum light, is emitted bluewards of 3300~\AA.
The absence of excess luminosity at late times, the cutoff of the spectral
energy distribution bluewards of 3000~\AA, and the absence of narrow line
emission and strong \ion{Na}{1}~D absorption all argue against a significant
contribution from ongoing shock interaction.  However, 
$\sim 10$\% of LSQ12gdj's luminosity near maximum light could be
produced by the release of trapped radiation, including kinetic energy
thermalized during a brief interaction with a compact, hydrogen-poor envelope
(radius $< 10^{13}$~cm) shortly after explosion; such an envelope arises
generically in double-degenerate merger scenarios.
\end{abstract}



\vspace{0.1in}

\section{Introduction}

Type Ia supernovae (SNe~Ia) have become indispensable as luminosity
distance indicators at large distances appropriate for studying the
cosmological dark energy \citep{riess98,scp99}.  They are believed to be the
thermonuclear explosions of carbon-oxygen white dwarfs, and their spectra are
generally very similar near maximum light, although some spectroscopic
diversity exists \citep{bfn93,benetti05,branch06,branch07,branch08,wang09}.

SNe~Ia used for cosmology are referred to as spectroscopically
``(Branch) normal'' \citep{bfn93} SNe~Ia;
they have a typical absolute magnitude near maximum light in the range
$-18.5 < M_V < -19.5$.  They are used as robust standard candles based on
empirical relations between the SN's luminosity and its colour
and light curve width \citep{riess96,tripp98,phillips99,goldhaber01}.
Maximum-light spectroscopic properties can also help to improve the precision
of distances measured using normal SNe~Ia \citep{sjb09,wang09,csp10,fk11}.

Another subclass of SNe~Ia with absolute magnitude $M_V \sim -20$ has also
attracted recent attention.  At least three events are currently known:
SN~2003fg \citep{howell06}, SN~2007if \citep{scalzo10,yuan10}, and SN~2009dc
\citep{yamanaka09,tanaka10,silverman11,taub11}.  A fourth event, SN~2006gz
\citep{hicken07}, is usually classed with these three, although its
maximum-light luminosity depends on an uncertain extinction correction from
dust in its host galaxy.  These four events are spectroscopically very
different from each other.  SN~2006gz has a photospheric velocity typical of
normal SNe~Ia as inferred from the velocity of the \ion{Si}{2} $\lambda 6355$
absorption minimum, and shows \ion{C}{2} absorption
($\lambda\lambda 4745, 6580, 7234$) in spectra taken more than 10~days before
$B$-band maximum light.  In contrast, SN~2009dc shows low \ion{Si}{2} velocity
\vSi\ ($\sim 8000$~\kms), a relatively high \ion{Si}{2} velocity gradient
\vdot\ ($\sim -75$~\kms~day$^{-1}$),
and very strong, persistent \ion{C}{2} $\lambda6580$ absorption.
SN~2007if is spectroscopically similar to SN~1991T
\citep{filippenko92,phillips92} before maximum light, its spectrum
dominated by \ion{Fe}{3} and showing only very weak \ion{Si}{2}, with a
definite \ion{C}{2} detection in a spectrum taken 5~days after $B$-band
maximum light.  SN~2006gz, SN~2007if and SN~2009dc show low-ionization
nebular spectra dominated by \ion{Fe}{2}, in contrast to normal SNe~Ia
which have stronger \ion{Fe}{3} emission \citep{maeda09,taub13}.  Only one
spectrum, taken at 2~days past $B$-band maximum, exists for SN~2003fg,
which resembles SN~2009dc at a similar phase.  Recently two additional SNe,
SN~2011aa and SN~2012dn, have been proposed as super-Chandrasekhar-mass
SN~Ia candidates based on their luminosity at ultraviolet (UV) wavelengths
as observed with the \emph{Swift} telescope \citep{brown14}.

These extremely luminous SNe~Ia cannot presently be explained by models of
exploding Chandrasekhar-mass white dwarfs, since the latter produce at most
1~\Msol\ of \nickel\ even in a pure detonation \citep{kmh93}.
While they might more descriptively be called ``superluminous SNe~Ia'',
these SNe~Ia have typically been referred to as
``candidate super-Chandrasekhar SNe~Ia'' or ``super-Chandras'', based on an
early interpretation of SN~2003fg as arising from the explosion of a
differentially rotating white dwarf with mass $\sim 2$~\Msol\
\citep{howell06}.  Observation of events
in this class has stimulated much recent theoretical investigation
into super-Chandrasekhar-mass SN~Ia channels
\citep{hachisu11,justham11,rds12,dm13a,dm13b}, and into mechanisms for
increasing the peak luminosity of Chandrasekhar-mass events \citep{hsr07}.

The status of superluminous SNe~Ia as being super-Chandrasekhar-mass has
historically been closely tied to their peak luminosity.  SN~2003fg's ejected
mass was inferred at first from its peak absolute magnitude $M_V = -19.94$,
requiring a large mass of \nickel\
\citep[$M_\mathrm{Ni} = 1.3 \pm 0.1$~\Msol;][]{arnett82}
and a low \ion{Si}{2} velocity near maximum ($\sim 8000$~\kms), suggesting
a high binding energy for the progenitor.  Ejected mass estimates were later
made for SN~2007if \citep{scalzo10} and SN~2009dc \citep{silverman11,taub11},
producing numbers of similar magnitude.  These ejected mass estimates depend,
to varying extents, on the interpretation of the maximum-light luminosity in
terms of a large \nickel\ mass, which can be influenced by asymmetries and/or
non-radioactive sources of luminosity.  For example, shock interaction with
a dense shroud of circumstellar material (CSM) has been proposed as a source
of luminosity near maximum light for SN~2009dc
\citep{taub11,hachinger12,taub13}.
The CSM envelope would have to be largely free of hydrogen and helium to
avoid producing emission lines of these elements in the shocked material.
The additional luminosity could simply represent trapped radiation from a
short interaction soon after explosion with a compact envelope, rather than
an ongoing interaction with an extended wind.  Such an envelope is naturally
produced in an explosion resulting from a ``slow'' merger of
two carbon-oxygen white dwarfs \citep{it84,shen12}.  \citet{kmh93} modeled
detonations of carbon-oxygen white dwarfs inside compact envelopes, calling
them \emph{tamped detonations}; these events are luminous and have long rise
times, but appear much like normal SNe~Ia after maximum light.  A strong
ongoing interaction with an extended wind, in contrast, is expected to
produce very broad, ultraviolet (UV)-bright light curves and blue,
featureless spectra uncharacteristic of normal SNe~Ia \citep{fryer10,bs10}.

Searching for more candidate super-Chandrasekhar-mass SNe~Ia, \citet{scalzo12}
reconstructed masses for a sample of SNe~Ia with spectroscopic behavior
matching a classical 1991T-like template and showing very slow evolution of
the \ion{Si}{2} velocity, similar to SN~2007if; these events were interpreted
as tamped detonations, and the mass reconstruction featured a very rough
accounting for trapped radiation.  One additional plausible
super-Chandrasekhar-mass candidate event was found, SNF~20080723-012,
with estimated ejected mass $\sim 1.7 \Msol$ and \nickel\ mass
$\sim 0.8 \Msol$.  The other events either had insufficient data to establish
super-Chandrasekhar-mass status with high confidence, or had reconstructed
masses consistent with the Chandrasekhar mass.  However, none of the
\citet{scalzo12} SNe had coverage at wavelengths bluer than 3300~\AA,
making it impossible to search for early signatures of shock interaction,
and potentially underestimating the maximum bolometric luminosity and the
\nickel\ mass.  While \citet{brown14} obtained good UV coverage of two new
candidate super-Chandrasekhar-mass SNe~Ia, 2011aa and 2012dn, no optical
photometry redward of 6000~\AA\ has yet been published for these SNe,
precluding the construction of their bolometric light curves
or detailed inference of their masses.

In this paper we present observations of a new overluminous ($M_B = -19.8$)
1991T-like SN~Ia, LSQ12gdj, including detailed UV (from \emph{Swift}) and
optical photometric coverage, as well as spectroscopic time series, starting
at 10~days before $B$-band maximum light.  We examine the UV behavior as
a tracer of shock interaction and as a contribution to the total bolometric
flux, \revised{and perform some simple semi-analytic modeling to address
the question:  what physical mechanisms can drive the high peak luminosity
in super-Chandrasekhar-mass SN~Ia candidates, and how might this relate to
the explosion mechanism(s) and the true progenitor mass?}


\section{Observations}
\label{sec:observations}


\subsection{Discovery and Classification}

LSQ12gdj was discovered on 2012~Nov~07 UT as part of the La Silla-QUEST
(LSQ) Low-Redshift Supernova Survey \citep{baltay13}, ongoing since 2009
using the QUEST-II camera mounted on the ESO 1-m Schmidt telescope at
La Silla Observatory.  QUEST-II observations are taken in a broad bandpass
using a custom-made interference filter with appreciable transmission
from 4000--7000~\AA, covering the SDSS $g'$ and $r'$ bandpasses.  Magnitudes
were calibrated in the LSQ natural system against stars in the SN field
with entries in the AAVSO All-Sky Photometric Survey (APASS) DR6 catalog.
These images are processed regularly using an image subtraction pipeline,
which uses reliable open-source software modules to subtract template
images of the constant night sky, leaving variable objects.
Each new image is registered and resampled to the position of a template
image using \code{SWarp} \citep{swarp}.  The template image is then rescaled
and convolved to match the point spread function (PSF) of the new image,
before being subtracted from the new image using \code{hotpants}\footnote{
http://www.astro.washington.edu/users/becker/hotpants.html}.
New objects on the subtracted images are detected using \code{SExtractor}
\citep{sextractor}.  These candidates are then visually scanned and the most
promising candidates selected for spectroscopic screening and follow-up.

\begin{figure}
\begin{center}
\resizebox{0.45\textwidth}{!}{\includegraphics{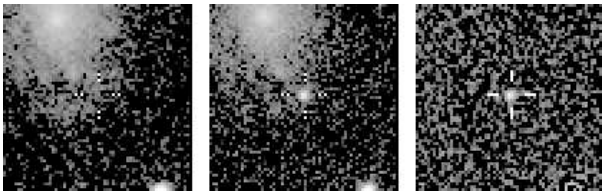}}
\caption{\small Discovery images for LSQ12gdj.
Left:  REF (galaxy template) image showing the host galaxy before the SN.
Center:  NEW image showing host galaxy + SN.  
Right:  Subtraction SUB = NEW - REF, showing the SN alone.
The thumbnails are $56'' \times 56''$ square.}
\label{fig:lsq-thumb}
\end{center}
\end{figure}

The discovery image of LSQ12gdj, showing its position
(RA = 23:54:43.32, DEC = $-25$:40:34.0) on the outskirts
of its host galaxy, ESO~472-~G~007 \citep[$z = 0.030324$;][]{z12gdj},
is shown in Figure~\ref{fig:lsq-thumb},
along with the galaxy template image and the subtracted image.
No source was detected at the SN position two days earlier (2012~Nov~05 UT)
to a limiting magnitude of $\sim 21$.
Ongoing LSQ observations of LSQ12gdj were taken after discovery as part of
the LSQ rolling search strategy, characterizing the rising part of the
light curve.  The early light curve of LSQ12gdj is shown in
Table~\ref{tbl:questphot}.

The Nearby Supernova Factory (SNfactory) reported that a spectrum taken
2013 Nov 10.2 UT with the SuperNova Integral Field Spectrograph
\citep[SNIFS;][]{snifs}
on the University of Hawaii 2.2-m telescope was a good match to a
1991T-like SN~Ia before maximum light as classified using SNID \citep{snid},
and flagged it as a candidate super-Chandrasekhar-mass SN~Ia \citep{atel-snf}.
This classification was later confirmed by the first PESSTO spectrum
in the time series described below, taken 2012 Nov 13 UT.


\subsection{Photometry}

\begin{figure}
\begin{center}
\resizebox{0.5\textwidth}{!}{\includegraphics{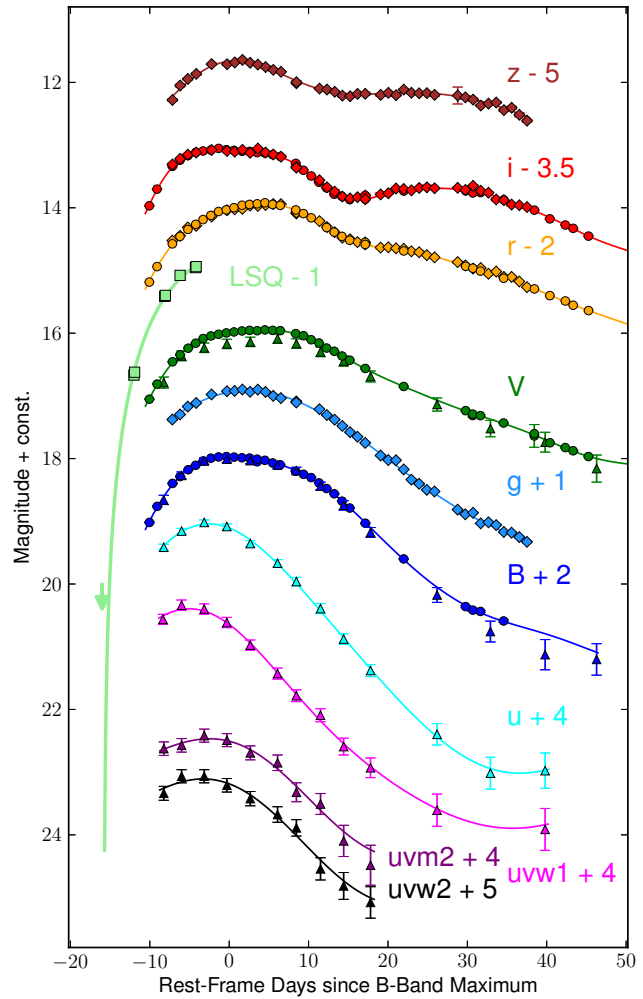}}
\caption{Multi-band light curves of LSQ12gdj.  CSP and LCOGT points are shown
as circles, \emph{Swift} points are upward-facing triangles, and LSQ points
are squares.  Solid curves:  Gaussian process regression fit to the data
\citep[see][]{rw06,scalzo14a}, except LSQ, for which the \citet{arnett82}
functional form is used (well-approximated by a $t^2$ rise at early times).
Light curve phase is with respect to $B$-band maximum
at MJD 56252.5 (2012 Nov 21.5).}
\label{fig:multi-lc}
\end{center}
\end{figure}

\emph{Swift} UVOT observations were triggered immediately after spectroscopic
confirmation, providing comprehensive photometric coverage at UV wavelengths
starting 8~days before $B$-band maximum light.  The observations were reduced
using aperture photometry according to the procedure in \citet{brown09},
using the updated zeropoints, sensitivity corrections, and transmission
curves of \citet{breeveld11}.

Ground-based follow-up photometry was taken by the
\emph{Carnegie Supernova Project II}
(CSP) using the Swope 1-m telescope at Las Campanas observatory,
in the natural system CSP $BVr'i'$ filters, starting at 10~days before
$B$-band maximum light.  The SITe3 CCD detector mounted on the Swope has
a 2048~$\times$~4096~pix active area, with a pixel scale of 0.435~arcsec/pix;
to reduce readout time, a 1200~$\times$~1200~pix subraster is read out,
for a field of view of $8.7 \times 8.7$ arcmin.
The images were reduced with standard CSP software including bias subtraction,
linearity correction, flat fielding and exposure correction.
A local sequence of 20 stars around the SN, covering a wide range of
magnitudes, has been calibrated on more than 15 photometric nights into the
natural system of the Swope telescope, using the reduction procedures
described in \citet{contreras10} and the bandpass calibration procedures and
transmission functions in \citet{cspdr2}.  Template images for galaxy
subtractions were taken with the Du~Pont 2.5-m telescope under favorable
seeing conditions on the nights of 2013~Oct~10--11, using the same filter
set as the science images.  PSF-fitting
photometry was performed on the SN detections in the template-subtracted
images, relative to the local sequence stars, measured with the standard IRAF
\citep{iraf} package \code{daophot} \citep{daophot}.

Additional ground-based photometry was taken by the Las Cumbres Observatory
Global Telescope Network (LCOGT).  The LCOGT data were reduced using a custom
pipeline developed by the LCOGT SN team, using standard procedures
(\code{pyraf}, \code{daophot}, \code{SWarp}) in a \code{python} framework.
PSF-fitting photometry is performed after subtraction of the background,
estimated via a low-order polynomial fit.

The \emph{Swift} UV photometry and the CSP/LCOGT optical photometry are shown
in Tables~\ref{tbl:photometry-swift} and \ref{tbl:photometry-opt},
respectively, and plotted in Figure~\ref{fig:multi-lc}.  All ground-based
magnitudes have been $S$-corrected to the appropriate standard system
\cite{landoltsys,sdsssys}.  The natural-system CSP/LCOGT photometry is shown
in Table~\ref{tbl:natphot} (online-only).

\begin{table}
\center
\caption{Ground-based photometry of LSQ12gdj in instrumental fluxes}
\begin{tabular}{rrrr}
\hline 
MJD & Phase$^a$ & Flux (ADU) & Inst. Mag$^b$ \\
\hline 
56238.074 & $-14.0$ & $ 1610.9 \pm  38.2$ & $18.98 \pm 0.03$ \\
56238.158 & $-13.9$ & $ 1751.2 \pm  44.2$ & $18.89 \pm 0.03$ \\
56240.061 & $-12.1$ & $ 5407.8 \pm  68.8$ & $17.67 \pm 0.01$ \\
56240.144 & $-12.0$ & $ 5637.9 \pm  77.6$ & $17.62 \pm 0.01$ \\
56244.054 & $ -8.2$ & $17167.6 \pm 101.1$ & $16.41 \pm 0.01$ \\
56244.138 & $ -8.1$ & $17428.5 \pm 108.0$ & $16.40 \pm 0.01$ \\
56246.049 & $ -6.3$ & $23382.5 \pm 186.7$ & $16.08 \pm 0.01$ \\
56248.044 & $ -4.3$ & $26198.3 \pm 230.2$ & $15.95 \pm 0.01$ \\
56248.128 & $ -4.2$ & $26575.1 \pm 236.4$ & $15.94 \pm 0.01$ \\
\hline 
\end{tabular}
\medskip \\
\flushleft
$^a$ Phase given in rest-frame days since $B$-band maximum light.\\
$^b$ Instrumental magnitudes assume a zeropoint of 27.0.  Upper limits are 95\% CL.
     Error bars are statistical only; systematic error is about 5\%.

\label{tbl:questphot}
\end{table}

\begin{table*}
\center
\caption{Swift photometry of LSQ12gdj}
\begin{tabular}{rrrrrrrr}
\hline 
MJD & Phase$^a$ & $uvw2$ & $uvm2$ & $uvw1$ & $u$ & $b$ & $v$ \\
\hline 
56243.9 & $-8.4$ & $18.34 \pm 0.11$ & $18.63 \pm 0.11$ & $16.56 \pm 0.08$ & $15.41 \pm 0.05$ & $16.65 \pm 0.07$ & $16.77 \pm 0.09$ \\
56246.2 & $-6.1$ & $18.07 \pm 0.10$ & $18.57 \pm 0.11$ & $16.34 \pm 0.08$ & $15.16 \pm 0.04$ & $16.25 \pm 0.06$ & $16.34 \pm 0.08$ \\
56249.2 & $-3.2$ & $18.07 \pm 0.10$ & $18.42 \pm 0.10$ & $16.40 \pm 0.08$ & $15.02 \pm 0.04$ & $16.02 \pm 0.05$ & $16.21 \pm 0.08$ \\
56252.1 & $-0.4$ & $18.21 \pm 0.11$ & $18.49 \pm 0.10$ & $16.61 \pm 0.08$ & $15.08 \pm 0.04$ & $15.98 \pm 0.05$ & $16.15 \pm 0.07$ \\
56255.2 & $ 2.6$ & $18.42 \pm 0.11$ & $18.69 \pm 0.11$ & $16.98 \pm 0.08$ & $15.36 \pm 0.05$ & $16.00 \pm 0.05$ & $16.12 \pm 0.07$ \\
56258.7 & $ 6.0$ & $18.68 \pm 0.12$ & $18.85 \pm 0.12$ & $17.43 \pm 0.09$ & $15.67 \pm 0.06$ & $16.08 \pm 0.05$ & $16.07 \pm 0.07$ \\
56261.1 & $ 8.4$ & $18.89 \pm 0.13$ & $19.32 \pm 0.15$ & $17.78 \pm 0.09$ & $15.96 \pm 0.07$ & $16.17 \pm 0.05$ & $16.14 \pm 0.07$ \\
56264.3 & $11.4$ & $19.54 \pm 0.18$ & $19.51 \pm 0.17$ & $18.09 \pm 0.10$ & $16.39 \pm 0.08$ & $16.42 \pm 0.06$ & $16.27 \pm 0.08$ \\
56267.3 & $14.3$ & $19.82 \pm 0.21$ & $20.10 \pm 0.25$ & $18.59 \pm 0.13$ & $16.88 \pm 0.09$ & $16.73 \pm 0.08$ & $16.42 \pm 0.08$ \\
56270.8 & $17.7$ & $20.08 \pm 0.25$ & $20.48 \pm 0.32$ & $18.93 \pm 0.16$ & $17.38 \pm 0.10$ & $17.18 \pm 0.08$ & $16.66 \pm 0.09$ \\
56279.4 & $26.1$ &           \ldots &           \ldots & $19.61 \pm 0.26$ & $18.40 \pm 0.17$ & $18.18 \pm 0.12$ & $17.11 \pm 0.10$ \\
56286.3 & $32.8$ &           \ldots &           \ldots &           \ldots & $19.02 \pm 0.25$ & $18.76 \pm 0.17$ & $17.49 \pm 0.13$ \\
56293.4 & $39.7$ &           \ldots &           \ldots & $19.91 \pm 0.33$ & $18.98 \pm 0.28$ & $19.11 \pm 0.24$ & $17.71 \pm 0.16$ \\
56300.1 & $46.2$ &           \ldots &           \ldots &           \ldots &           \ldots & $19.17 \pm 0.25$ & $18.13 \pm 0.22$ \\
\hline 
\end{tabular}
\medskip \\
\flushleft
$^a$ Phase given in rest-frame days since $B$-band maximum light.\\

\label{tbl:photometry-swift}
\end{table*}

\begin{table*}
\center
\caption{Ground-based photometry of LSQ12gdj in the Landolt and SDSS
         standard systems}
\begin{tabular}{rrrrrrrrr}
\hline 
MJD & Phase$^a$ & $B$ & $V$ & $g$ & $r$ & $i$ & $z$ & Source\\
\hline 
56242.1 & $-10.2$ & $17.02 \pm 0.01$ & $17.05 \pm 0.01$ &           \ldots & $17.19 \pm 0.01$ & $17.47 \pm 0.01$ &           \ldots & SWOPE \\
56243.1 & $ -9.2$ & $16.76 \pm 0.01$ & $16.81 \pm 0.01$ &           \ldots & $16.94 \pm 0.01$ & $17.20 \pm 0.01$ &           \ldots & SWOPE \\
56245.0 & $ -7.3$ &           \ldots &           \ldots & $16.38 \pm 0.01$ & $16.52 \pm 0.01$ & $16.81 \pm 0.02$ & $17.28 \pm 0.01$ & LCOGT \\
56245.1 & $ -7.2$ & $16.40 \pm 0.01$ & $16.46 \pm 0.01$ &           \ldots & $16.57 \pm 0.01$ & $16.84 \pm 0.01$ &           \ldots & SWOPE \\
56246.0 & $ -6.3$ & $16.29 \pm 0.01$ & $16.34 \pm 0.01$ &           \ldots & $16.46 \pm 0.01$ & $16.74 \pm 0.01$ &           \ldots & SWOPE \\
56246.0 & $ -6.3$ &           \ldots &           \ldots & $16.30 \pm 0.01$ & $16.45 \pm 0.02$ & $16.72 \pm 0.01$ & $17.05 \pm 0.02$ & LCOGT \\
56247.0 & $ -5.3$ &           \ldots &           \ldots & $16.17 \pm 0.01$ & $16.31 \pm 0.02$ & $16.65 \pm 0.01$ & $16.95 \pm 0.02$ & LCOGT \\
56247.1 & $ -5.3$ & $16.18 \pm 0.01$ & $16.24 \pm 0.01$ &           \ldots & $16.35 \pm 0.01$ & $16.67 \pm 0.01$ &           \ldots & SWOPE \\
56248.1 & $ -4.3$ & $16.11 \pm 0.01$ & $16.16 \pm 0.01$ &           \ldots & $16.26 \pm 0.00$ & $16.62 \pm 0.01$ &           \ldots & SWOPE \\
56248.1 & $ -4.3$ &           \ldots &           \ldots & $16.11 \pm 0.01$ & $16.29 \pm 0.01$ & $16.62 \pm 0.02$ & $16.86 \pm 0.02$ & LCOGT \\
56249.0 & $ -3.4$ & $16.04 \pm 0.01$ & $16.09 \pm 0.01$ &           \ldots & $16.19 \pm 0.00$ & $16.59 \pm 0.01$ &           \ldots & SWOPE \\
56250.0 & $ -2.4$ & $15.99 \pm 0.01$ & $16.04 \pm 0.01$ &           \ldots & $16.12 \pm 0.01$ & $16.58 \pm 0.01$ &           \ldots & SWOPE \\
56250.1 & $ -2.3$ &           \ldots &           \ldots & $15.98 \pm 0.01$ & $16.14 \pm 0.02$ & $16.58 \pm 0.01$ & $16.71 \pm 0.02$ & LCOGT \\
56251.0 & $ -1.4$ & $15.97 \pm 0.01$ & $16.02 \pm 0.01$ &           \ldots & $16.06 \pm 0.01$ & $16.55 \pm 0.01$ &           \ldots & SWOPE \\
56252.0 & $ -0.5$ & $15.97 \pm 0.01$ & $16.00 \pm 0.01$ &           \ldots & $16.03 \pm 0.01$ & $16.57 \pm 0.01$ &           \ldots & SWOPE \\
56252.1 & $ -0.4$ &           \ldots &           \ldots & $15.93 \pm 0.01$ & $16.04 \pm 0.01$ & $16.59 \pm 0.01$ & $16.71 \pm 0.01$ & LCOGT \\
56253.0 & $  0.5$ & $15.98 \pm 0.01$ & $15.98 \pm 0.01$ &           \ldots & $16.00 \pm 0.00$ & $16.58 \pm 0.01$ &           \ldots & SWOPE \\
56253.1 & $  0.6$ &           \ldots &           \ldots & $15.92 \pm 0.01$ & $16.03 \pm 0.01$ & $16.61 \pm 0.01$ & $16.68 \pm 0.01$ & LCOGT \\
56254.1 & $  1.5$ & $15.99 \pm 0.01$ & $15.96 \pm 0.01$ &           \ldots & $15.96 \pm 0.01$ & $16.60 \pm 0.01$ &           \ldots & SWOPE \\
56254.1 & $  1.6$ &           \ldots &           \ldots & $15.90 \pm 0.01$ & $16.02 \pm 0.01$ & $16.58 \pm 0.02$ & $16.64 \pm 0.01$ & LCOGT \\
56255.1 & $  2.5$ & $15.99 \pm 0.01$ & $15.96 \pm 0.01$ &           \ldots & $15.95 \pm 0.01$ & $16.60 \pm 0.01$ &           \ldots & SWOPE \\
56255.1 & $  2.5$ &           \ldots &           \ldots & $15.93 \pm 0.01$ & $15.97 \pm 0.01$ & $16.62 \pm 0.01$ & $16.69 \pm 0.02$ & LCOGT \\
56256.1 & $  3.4$ & $16.05 \pm 0.02$ & $15.96 \pm 0.01$ &           \ldots & $15.94 \pm 0.01$ & $16.62 \pm 0.01$ &           \ldots & SWOPE \\
56256.1 & $  3.5$ &           \ldots &           \ldots & $15.90 \pm 0.01$ & $15.97 \pm 0.01$ & $16.55 \pm 0.02$ & $16.72 \pm 0.02$ & LCOGT \\
56257.1 & $  4.4$ & $16.03 \pm 0.01$ & $15.95 \pm 0.01$ &           \ldots & $15.92 \pm 0.01$ & $16.63 \pm 0.01$ &           \ldots & SWOPE \\
56257.1 & $  4.5$ &           \ldots &           \ldots & $15.94 \pm 0.01$ & $15.96 \pm 0.01$ & $16.62 \pm 0.02$ & $16.76 \pm 0.02$ & LCOGT \\
56258.1 & $  5.4$ & $16.07 \pm 0.01$ & $15.95 \pm 0.01$ &           \ldots & $15.95 \pm 0.01$ & $16.65 \pm 0.01$ &           \ldots & SWOPE \\
56258.1 & $  5.5$ &           \ldots &           \ldots & $16.00 \pm 0.01$ & $15.94 \pm 0.01$ & $16.65 \pm 0.01$ & $16.80 \pm 0.02$ & LCOGT \\
56259.0 & $  6.3$ & $16.10 \pm 0.01$ & $15.96 \pm 0.01$ &           \ldots & $15.96 \pm 0.01$ & $16.68 \pm 0.01$ &           \ldots & SWOPE \\
56259.1 & $  6.4$ &           \ldots &           \ldots & $16.03 \pm 0.01$ & $15.94 \pm 0.01$ & $16.69 \pm 0.03$ & $16.83 \pm 0.02$ & LCOGT \\
56261.1 & $  8.3$ & $16.20 \pm 0.01$ & $16.01 \pm 0.01$ &           \ldots & $16.03 \pm 0.01$ & $16.79 \pm 0.01$ &           \ldots & SWOPE \\
56261.1 & $  8.4$ &           \ldots &           \ldots & $16.09 \pm 0.01$ & $16.09 \pm 0.01$ & $16.85 \pm 0.01$ & $17.01 \pm 0.01$ & LCOGT \\
56262.1 & $  9.3$ & $16.25 \pm 0.01$ & $16.04 \pm 0.01$ &           \ldots & $16.09 \pm 0.01$ & $16.90 \pm 0.01$ &           \ldots & SWOPE \\
56263.1 & $ 10.2$ & $16.30 \pm 0.01$ & $16.10 \pm 0.01$ &           \ldots & $16.16 \pm 0.01$ & $16.99 \pm 0.01$ &           \ldots & SWOPE \\
56264.1 & $ 11.2$ & $16.39 \pm 0.01$ & $16.16 \pm 0.01$ &           \ldots &           \ldots & $17.09 \pm 0.01$ &           \ldots & SWOPE \\
56264.1 & $ 11.3$ &           \ldots &           \ldots & $16.21 \pm 0.01$ & $16.21 \pm 0.03$ & $17.06 \pm 0.00$ & $17.10 \pm 0.02$ & LCOGT \\
\hline 
\end{tabular}
\medskip \\
\flushleft
$^a$ Phase given in rest-frame days since $B$-band maximum light.\\

\label{tbl:photometry-opt}
\end{table*}

\begin{table*}
\addtocounter{table}{-1}
\center
\caption{Ground-based photometry of LSQ12gdj, cont'd.}
\begin{tabular}{rrrrrrrrr}
\hline 
MJD & Phase$^a$ & $B$ & $V$ & $g$ & $r$ & $i$ & $z$ & Source\\
\hline 
56265.1 & $ 12.2$ & $16.48 \pm 0.01$ & $16.24 \pm 0.01$ &           \ldots & $16.31 \pm 0.01$ & $17.18 \pm 0.01$ &           \ldots & SWOPE \\
56265.1 & $ 12.2$ &           \ldots &           \ldots & $16.31 \pm 0.01$ & $16.31 \pm 0.01$ & $17.17 \pm 0.01$ & $17.11 \pm 0.01$ & LCOGT \\
56266.1 & $ 13.1$ & $16.56 \pm 0.01$ & $16.29 \pm 0.01$ &           \ldots & $16.39 \pm 0.01$ & $17.28 \pm 0.01$ &           \ldots & SWOPE \\
56266.1 & $ 13.2$ &           \ldots &           \ldots & $16.38 \pm 0.01$ & $16.39 \pm 0.01$ & $17.24 \pm 0.03$ & $17.15 \pm 0.02$ & LCOGT \\
56267.1 & $ 14.1$ & $16.68 \pm 0.01$ & $16.36 \pm 0.01$ &           \ldots & $16.46 \pm 0.01$ & $17.32 \pm 0.01$ &           \ldots & SWOPE \\
56267.1 & $ 14.2$ &           \ldots &           \ldots & $16.48 \pm 0.01$ & $16.48 \pm 0.01$ & $17.32 \pm 0.01$ & $17.21 \pm 0.02$ & LCOGT \\
56268.1 & $ 15.1$ & $16.79 \pm 0.01$ & $16.43 \pm 0.01$ &           \ldots & $16.52 \pm 0.01$ & $17.37 \pm 0.01$ &           \ldots & SWOPE \\
56268.1 & $ 15.1$ &           \ldots &           \ldots & $16.58 \pm 0.01$ & $16.50 \pm 0.01$ & $17.35 \pm 0.01$ & $17.22 \pm 0.02$ & LCOGT \\
56269.1 & $ 16.1$ &           \ldots &           \ldots & $16.65 \pm 0.01$ & $16.55 \pm 0.01$ & $17.33 \pm 0.01$ & $17.18 \pm 0.02$ & LCOGT \\
56270.1 & $ 17.0$ &           \ldots &           \ldots & $16.75 \pm 0.01$ & $16.55 \pm 0.01$ & $17.30 \pm 0.02$ & $17.18 \pm 0.02$ & LCOGT \\
56270.1 & $ 17.0$ & $17.03 \pm 0.01$ & $16.56 \pm 0.01$ &           \ldots & $16.58 \pm 0.01$ & $17.36 \pm 0.01$ &           \ldots & SWOPE \\
56272.1 & $ 19.0$ &           \ldots &           \ldots & $16.95 \pm 0.01$ & $16.63 \pm 0.02$ & $17.29 \pm 0.03$ & $17.20 \pm 0.02$ & LCOGT \\
56273.1 & $ 20.0$ &           \ldots &           \ldots & $17.02 \pm 0.01$ & $16.63 \pm 0.01$ & $17.27 \pm 0.01$ & $17.17 \pm 0.01$ & LCOGT \\
56274.1 & $ 20.9$ &           \ldots &           \ldots & $17.03 \pm 0.01$ & $16.65 \pm 0.01$ & $17.19 \pm 0.01$ & $17.20 \pm 0.02$ & LCOGT \\
56275.1 & $ 21.9$ & $17.60 \pm 0.01$ & $16.85 \pm 0.01$ &           \ldots & $16.69 \pm 0.01$ & $17.23 \pm 0.01$ &           \ldots & SWOPE \\
56275.1 & $ 21.9$ &           \ldots &           \ldots & $17.17 \pm 0.01$ & $16.65 \pm 0.02$ & $17.20 \pm 0.01$ & $17.12 \pm 0.01$ & LCOGT \\
56276.1 & $ 22.9$ &           \ldots &           \ldots & $17.34 \pm 0.01$ & $16.69 \pm 0.01$ & $17.22 \pm 0.01$ & $17.17 \pm 0.01$ & LCOGT \\
56277.1 & $ 23.8$ &           \ldots &           \ldots & $17.40 \pm 0.01$ & $16.71 \pm 0.03$ & $17.18 \pm 0.01$ & $17.17 \pm 0.02$ & LCOGT \\
56278.1 & $ 24.8$ &           \ldots &           \ldots & $17.50 \pm 0.01$ & $16.76 \pm 0.01$ & $17.17 \pm 0.03$ & $17.17 \pm 0.01$ & LCOGT \\
56279.1 & $ 25.8$ &           \ldots &           \ldots & $17.52 \pm 0.01$ & $16.79 \pm 0.01$ & $17.19 \pm 0.01$ & $17.20 \pm 0.02$ & LCOGT \\
56282.1 & $ 28.7$ &           \ldots &           \ldots & $17.81 \pm 0.01$ & $16.87 \pm 0.02$ & $17.18 \pm 0.01$ & $17.21 \pm 0.14$ & LCOGT \\
56283.1 & $ 29.7$ &           \ldots &           \ldots & $17.89 \pm 0.01$ & $16.92 \pm 0.01$ & $17.22 \pm 0.01$ & $17.24 \pm 0.03$ & LCOGT \\
56283.1 & $ 29.7$ & $18.36 \pm 0.02$ & $17.24 \pm 0.01$ &           \ldots & $16.93 \pm 0.01$ & $17.23 \pm 0.01$ &           \ldots & SWOPE \\
56284.0 & $ 30.6$ & $18.42 \pm 0.02$ & $17.31 \pm 0.02$ &           \ldots & $16.96 \pm 0.01$ & $17.26 \pm 0.01$ &           \ldots & SWOPE \\
56284.1 & $ 30.6$ &           \ldots &           \ldots & $17.86 \pm 0.02$ & $16.97 \pm 0.01$ & $17.15 \pm 0.01$ & $17.28 \pm 0.02$ & LCOGT \\
56285.0 & $ 31.6$ & $18.44 \pm 0.02$ & $17.31 \pm 0.01$ &           \ldots & $17.01 \pm 0.01$ & $17.26 \pm 0.01$ &           \ldots & SWOPE \\
56285.1 & $ 31.6$ &           \ldots &           \ldots & $18.03 \pm 0.01$ & $16.96 \pm 0.01$ & $17.23 \pm 0.01$ & $17.37 \pm 0.02$ & LCOGT \\
56286.1 & $ 32.6$ &           \ldots &           \ldots & $18.02 \pm 0.01$ & $17.05 \pm 0.01$ & $17.26 \pm 0.01$ & $17.34 \pm 0.02$ & LCOGT \\
56287.1 & $ 33.5$ &           \ldots &           \ldots & $18.06 \pm 0.01$ & $17.02 \pm 0.02$ & $17.37 \pm 0.02$ & $17.32 \pm 0.01$ & LCOGT \\
56288.0 & $ 34.5$ & $18.59 \pm 0.02$ & $17.43 \pm 0.01$ &           \ldots & $17.12 \pm 0.01$ & $17.37 \pm 0.01$ &           \ldots & SWOPE \\
56288.1 & $ 34.5$ &           \ldots &           \ldots & $18.17 \pm 0.01$ & $17.10 \pm 0.01$ & $17.38 \pm 0.01$ & $17.44 \pm 0.02$ & LCOGT \\
56289.1 & $ 35.5$ &           \ldots &           \ldots & $18.18 \pm 0.02$ & $17.10 \pm 0.02$ & $17.45 \pm 0.03$ & $17.41 \pm 0.02$ & LCOGT \\
56290.1 & $ 36.4$ &           \ldots &           \ldots & $18.25 \pm 0.02$ & $17.18 \pm 0.01$ & $17.46 \pm 0.01$ & $17.52 \pm 0.03$ & LCOGT \\
56291.1 & $ 37.4$ &           \ldots &           \ldots & $18.33 \pm 0.01$ & $17.27 \pm 0.01$ & $17.49 \pm 0.01$ & $17.61 \pm 0.02$ & LCOGT \\
56292.0 & $ 38.4$ &           \ldots & $17.62 \pm 0.02$ &           \ldots & $17.29 \pm 0.01$ & $17.54 \pm 0.01$ &           \ldots & SWOPE \\
56294.0 & $ 40.3$ &           \ldots & $17.77 \pm 0.02$ &           \ldots & $17.40 \pm 0.01$ & $17.68 \pm 0.01$ &           \ldots & SWOPE \\
56296.1 & $ 42.2$ &           \ldots & $17.84 \pm 0.02$ &           \ldots & $17.49 \pm 0.01$ & $17.77 \pm 0.01$ &           \ldots & SWOPE \\
56297.1 & $ 43.2$ &           \ldots & $17.89 \pm 0.02$ &           \ldots & $17.55 \pm 0.01$ & $17.83 \pm 0.01$ &           \ldots & SWOPE \\
56299.1 & $ 45.2$ &           \ldots & $17.98 \pm 0.02$ &           \ldots & $17.64 \pm 0.01$ & $17.95 \pm 0.01$ &           \ldots & SWOPE \\
56316.0 & $ 61.6$ &           \ldots & $18.48 \pm 0.04$ &           \ldots & $18.31 \pm 0.03$ & $18.77 \pm 0.04$ &           \ldots & SWOPE \\
56318.0 & $ 63.6$ &           \ldots & $18.55 \pm 0.03$ &           \ldots &           \ldots &           \ldots &           \ldots & SWOPE \\
\hline 
\end{tabular}
\medskip \\
\flushleft
$^a$ Phase given in rest-frame days since $B$-band maximum light.\\

\end{table*}

\begin{table*}
\center
\caption{Ground-based photometry of LSQ12gdj in the natural systems of the
         Swope and LCOGT telescopes}
\begin{tabular}{rrrrrrrrr}
\hline 
MJD & Phase$^a$ & $B$ & $V$ & $g$ & $r$ & $i$ & $z$ & Source\\
\hline 
56242.1 & $-10.2$ & $17.02 \pm 0.01$ & $17.05 \pm 0.01$ &           \ldots & $17.20 \pm 0.01$ & $17.52 \pm 0.01$ &           \ldots & SWOPE \\
56243.1 & $ -9.2$ & $16.76 \pm 0.01$ & $16.81 \pm 0.01$ &           \ldots & $16.95 \pm 0.01$ & $17.25 \pm 0.01$ &           \ldots & SWOPE \\
56245.0 & $ -7.3$ &           \ldots &           \ldots & $16.39 \pm 0.01$ & $16.53 \pm 0.01$ & $16.83 \pm 0.02$ & $17.25 \pm 0.01$ & LCOGT \\
56245.1 & $ -7.2$ & $16.39 \pm 0.01$ & $16.46 \pm 0.01$ &           \ldots & $16.60 \pm 0.01$ & $16.88 \pm 0.01$ &           \ldots & SWOPE \\
56246.0 & $ -6.3$ & $16.28 \pm 0.01$ & $16.35 \pm 0.01$ &           \ldots & $16.48 \pm 0.01$ & $16.78 \pm 0.01$ &           \ldots & SWOPE \\
56246.0 & $ -6.3$ &           \ldots &           \ldots & $16.31 \pm 0.01$ & $16.46 \pm 0.02$ & $16.74 \pm 0.01$ & $17.02 \pm 0.02$ & LCOGT \\
56247.0 & $ -5.3$ &           \ldots &           \ldots & $16.18 \pm 0.01$ & $16.32 \pm 0.02$ & $16.67 \pm 0.01$ & $16.91 \pm 0.02$ & LCOGT \\
56247.1 & $ -5.3$ & $16.17 \pm 0.01$ & $16.24 \pm 0.01$ &           \ldots & $16.37 \pm 0.01$ & $16.71 \pm 0.01$ &           \ldots & SWOPE \\
\hline 
\end{tabular}
\medskip \\
\flushleft
The full version of this table can be found online. \\
$^a$ Phase given in rest-frame days since $B$-band maximum light.\\

\label{tbl:natphot}
\end{table*}


\subsection{Spectroscopy}

A full spectroscopic time series was taken by the Public ESO Spectroscopic
Survey for Transient Objects (PESSTO), using the EFOSC2 spectrograph
on the ESO New Technology Telescope (NTT) at La Silla Observatory,
comprising seven spectra taken between 2012 Nov 13 and 2013 Jan 13 UT.
The $gr11$ and $gr16$ gratings were used, covering the entire wavelength
range 3360--10330~\AA\ at 13~\AA\ resolution.  The spectra were reduced
using the \code{pyraf} package as part of a custom-built, Python-based
pipeline written for PESSTO; the pipeline includes corrections for bias and
fringing, wavelength and flux calibration, correction for telluric
absorption and a cross-check of the wavelength calibration using
atmospheric emission lines.

Three spectra of LSQ12gdj were obtained around maximum light by CSP using
the Las Campanas 2.5-m du Pont telescope and WFCCD.  The spectral resolution
is 8~\AA, as measured from the FWHM of the HeNeAr comparison lines.
A complete description of data reduction procedures
can be found in \citet{hamuy06}.

An additional five optical spectra were taken with the WiFeS integral field
spectrograph on the ANU 2.3-m telescope at Siding Spring Observatory.
WiFeS spectra were obtained using the B3000 and R3000 gratings, providing
wavelength coverage in the range 3500--9600~\AA\ with a FWHM resolution for
the point-spread function (PSF) of 1.5~\AA\ (blue channel) and 2.5 \AA\
(red channel).  Data cubes for WiFeS observations were produced using
the PyWiFeS\footnote{http://www.mso.anu.edu.au/pywifes/doku.php} software
\citep{pywifes}.  Spectra of the SN were extracted from final data cubes
using a PSF-weighted extraction technique with a simple symmetric Gaussian
PSF, and the width of this Gaussian was measured directly from the data cube.
Background subtraction was performed by calculating the median background
spectrum across all pixels outside a distance from the SN equal to about
three times the seeing disk (typically $1''.5$--$2''$ FWHM).
This technique produced good results for the WiFeS spectra of LSQ12gdj,
due to the negligible galaxy background and good spatial flatfielding
from the PyWiFeS pipeline.

\begin{figure*}
\begin{center}
\resizebox{\textwidth}{!}{\includegraphics{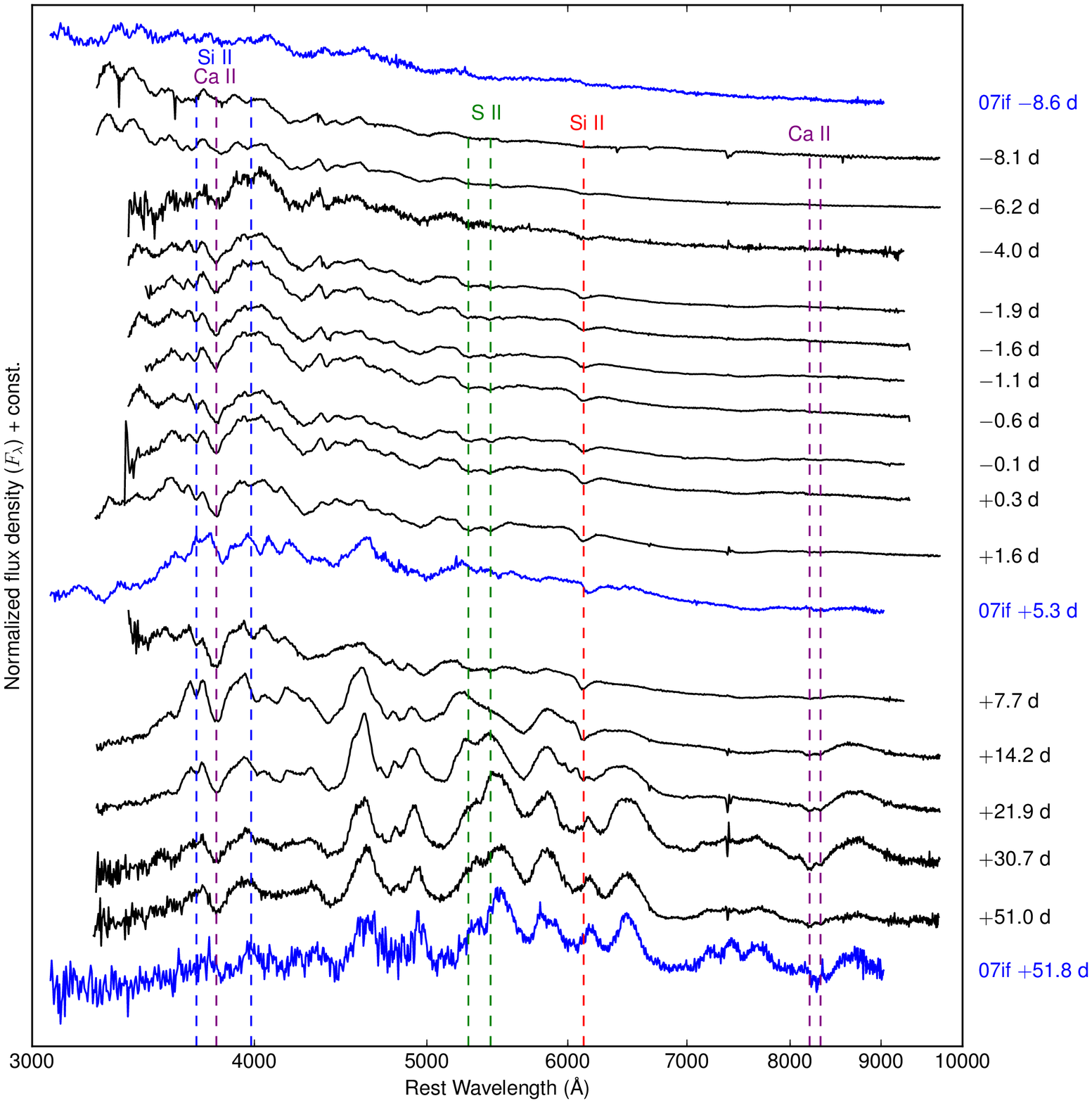}}
\caption{Spectral time series of LSQ12gdj
(black solid lines, phase labels on right), shown with spectra of SN~2007if
\citep[blue solid lines, rest-frame phase labels on right;][]{scalzo12}
for comparison.  All spectra have been rebinned to 5~\AA.  Constant
velocity locations for absorption minima of various intermediate-mass
element absorption features are marked with dashed lines:
purple, \ion{Ca}{2} H+K and NIR (10000~\kms);
blue, \ion{Si}{2} $\lambda\lambda 3858, 4129$ (11500~\kms);
red, \ion{Si}{2} $\lambda\lambda 6355$ (11000~\kms);
green, \ion{S}{2} $\lambda\lambda 5454, 5640$ (10500~\kms).}
\label{fig:spectra}
\end{center}
\end{figure*}

The observation log for all spectra presented is shown in
Table~\ref{tbl:spectroscopy}, and the spectra are plotted in
Figure~\ref{fig:spectra}.  All spectra will be publicly available
through WISeREP\footnote{http://www.weizmann.ac.il/astrophysics/wiserep/}
\citep{wiserep}.

\begin{table*}
\center
\caption{Optical spectroscopy of LSQ12gdj}
\begin{tabular}{lrrcrcl}
\hline
UT             & MJD     & Phase$^a$ & Telescope        & Exposure & Wavelength    & Observers$^b$ \\
Date           &         & (days)    & / Instrument     & Time (s) & Range (\AA)   &               \\
\hline
2012 Nov 13.13 & 56244.1 &  $-8.1$   & NTT-3.6m / EFOSC &            1500 & 3360--10000 & PESSTO \\
2012 Nov 15.14 & 56246.1 &  $-6.2$   & NTT-3.6m / EFOSC &            1500 & 3360--10000 & PESSTO \\
2012 Nov 17.43 & 56248.4 &  $-4.0$   & ANU-2.3m / WiFeS &            1200 & 3500--9550  & NS     \\
2012 Nov 19.52 & 56250.5 &  $-1.9$   & ANU-2.3m / WiFeS &            1200 & 3500--9550  & MC     \\
2012 Nov 19.92 & 56250.9 &  $-1.6$   & DuPont   / WFCCD & $2 \times  600$ & 3580--9620  & NM     \\
2012 Nov 20.43 & 56251.4 &  $-1.1$   & ANU-2.3m / WiFeS &            1200 & 3500--9550  & MC     \\
2012 Nov 20.93 & 56251.9 &  $-0.6$   & DuPont   / WFCCD & $2 \times  600$ & 3580--9620  & NM     \\
2012 Nov 21.45 & 56252.5 &  $-0.1$   & ANU-2.3m / WiFeS &            1200 & 3500--9550  & MC     \\
2012 Nov 21.85 & 56252.9 &  $+0.3$   & DuPont   / WFCCD & $2 \times  600$ & 3580--9620  & NM     \\
2012 Nov 23.15 & 56254.1 &  $+1.6$   & NTT-3.6m / EFOSC &             900 & 3360--10000 & PESSTO \\
2012 Nov 29.47 & 56260.5 &  $+7.7$   & ANU-2.3m / WiFeS &            1200 & 3500--9550  & CL,BS  \\
2012 Dec 06.12 & 56267.1 & $+14.2$   & NTT-3.6m / EFOSC &            1500 & 3360--10000 & PESSTO \\
2012 Dec 14.13 & 56275.1 & $+21.9$   & NTT-3.6m / EFOSC &            1500 & 3360--10000 & PESSTO \\
2012 Dec 23.13 & 56284.1 & $+30.7$   & NTT-3.6m / EFOSC &             900 & 3360--10000 & PESSTO \\
2013 Jan 13.05 & 56305.1 & $+51.0$   & NTT-3.6m / EFOSC & $2 \times 1500$ & 3360--10000 & PESSTO \\
\hline
\end{tabular}
\flushleft
$^a$ In rest-frame days with respect to B-band maximum (MJD 56252.4). \\
$^b$ 
BS = Brad Schaefer,
CL = Chris Lidman,
MC = Mike Childress,
NM = Nidia Morrell,
NS = Nicholas Scott

\label{tbl:spectroscopy}
\end{table*}

\section{Analysis}
\label{sec:analysis}

In this section we discuss quantities derived from the photometry and
spectroscopy in more detail.
We characterize the spectroscopic evolution of LSQ12gdj, including the
velocities of common absorption features, in \S\ref{subsec:spec}.
We discuss the broad-band light curves of LSQ12gdj and estimate the host
galaxy extinction in \S\ref{subsec:phot}.  Finally, we describe
the construction of a bolometric light curve for LSQ12gdj
in \S\ref{subsec:bolo}, including correction for unobserved NIR flux and
the process of solving for a low-resolution broad-band spectral energy
distribution (SED).


\subsection{Spectral Features and Velocity Evolution}
\label{subsec:spec}

Figure \ref{fig:spectra} shows the spectroscopic evolution of LSQ12gdj, with
spectra of the super-Chandrasekhar-mass SN~2007if included for comparison.
The early spectra show evidence for a hot photosphere, with a blue continuum
and absorption features dominated by \ion{Fe}{2} and \ion{Fe}{3}, typical of
1991T-like SNe~Ia \cite{filippenko92,phillips92}.
These include absorption complexes near 3500~\AA,
attributed to iron-peak elements (\ion{Ni}{2}, \ion{Co}{2}, and \ion{Cr}{2})
in SN~2007if \citep{scalzo10}.  The prominence of hot iron-peak elements
in the outer layers is consistent with a great deal of \nickel\ being
produced, and/or with significant mixing of \nickel\ throughout the outer
layers of ejecta during the explosion.  \ion{Si}{2}~$\lambda 5972$ is not
visible.  \ion{Si}{2}~$\lambda 6355$ and \ion{Ca}{2}~H+K are weak
throughout the evolution.

\begin{figure*}
\begin{center}
\resizebox{0.85\textwidth}{!}{\includegraphics{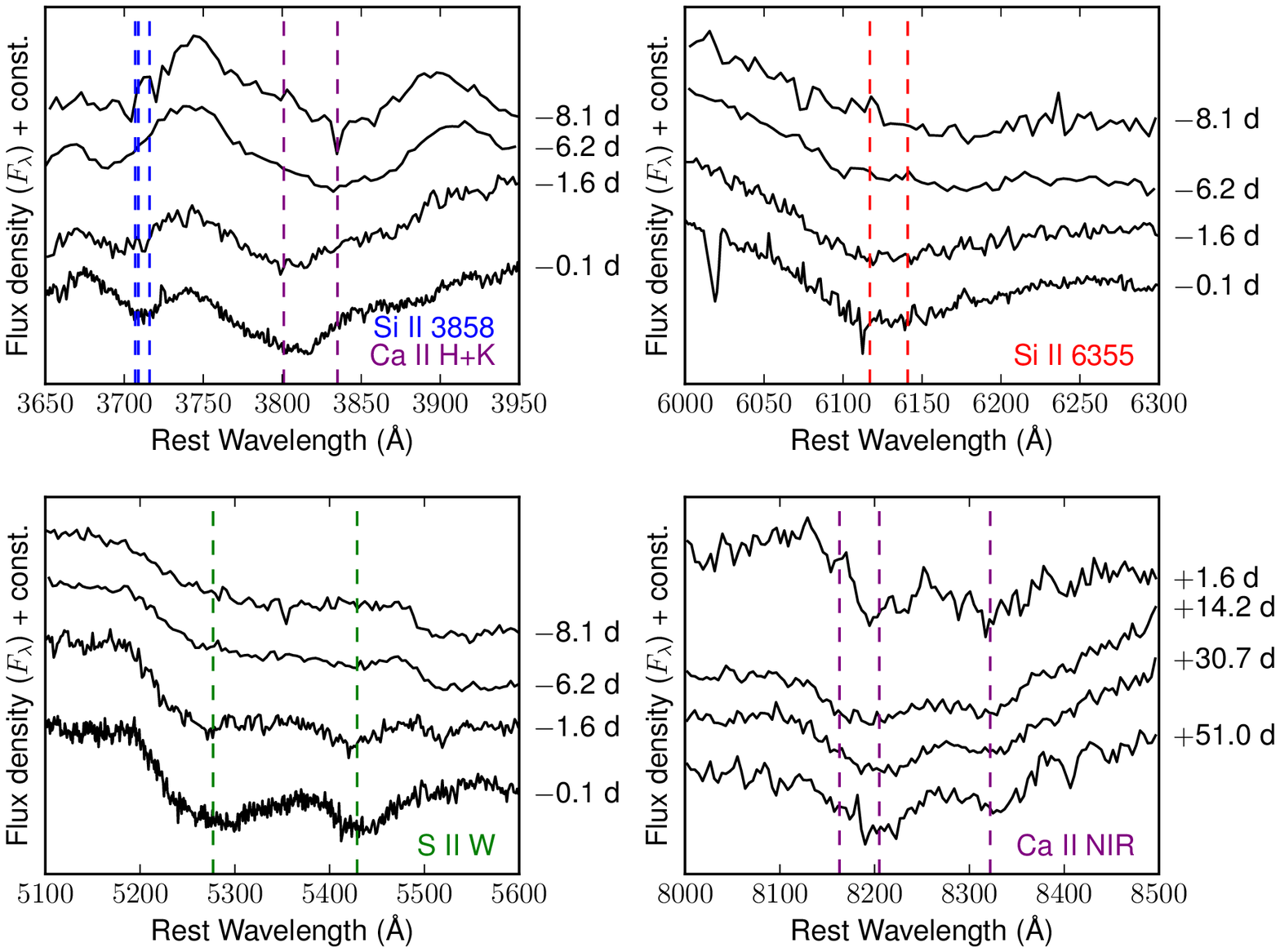}}
\caption{Subranges of spectra showing absorption line
profiles of intermediate-mass elements in the LSQ12gdj spectral time series.
Vertical dashed lines indicate the velocity of every component of
each absorbing multiplet near maximum light.  Phases shown along the
right-hand edges of the plots are in rest-frame days with respect to
$B$-band maximum light, as in Figure~\ref{fig:spectra}.}
\label{fig:linezoom}
\end{center}
\end{figure*}

Figure~\ref{fig:linezoom} shows subranges of the spectra highlighting
common intermediate-mass element lines at key points in their evolution.
LSQ12gdj shows unusually narrow intermediate-mass-element signatures.
The \ion{Ca}{2}~H+K absorption is narrow enough ($\sim 6000$~\kms FWHM)
that the minimum is unblended with neighboring \ion{Si}{2}~$\lambda3858$;
\revised{an inflection in the line profile redwards of the main minimum
could be signs that the doublet structure is just barely unresolved.
At later phases, the two reddest components of the \ion{Ca}{2} NIR triplet
show distinct minima near 12000~\kms.  The \ion{Si}{2}~$\lambda 6355$ line
profile near maximum light has a flat, boxy minimum.}
Spectra at the earliest phases show absorption minima near the expected
positions of all of these lines near maximum light, but with unexpected
shapes; these lines may not correspond physically to the nearest familiar
feature in each case, but if they do, they may yield interesting information
about the level populations to detailed modelling which properly
accounts for the ionization balance.  An example of such ambiguity is the
feature near 3650~\AA\ in the pre-maximum spectra, the position of which is
consistent with high-velocity \ion{Ca}{2} as in normal SNe~Ia, but is also
near the expected position of \ion{Si}{3} around 12000~\kms.

\begin{figure*}
\begin{center}
\resizebox{0.75\textwidth}{!}{\includegraphics
   [trim = 10mm 0mm 0mm 0mm, clip]{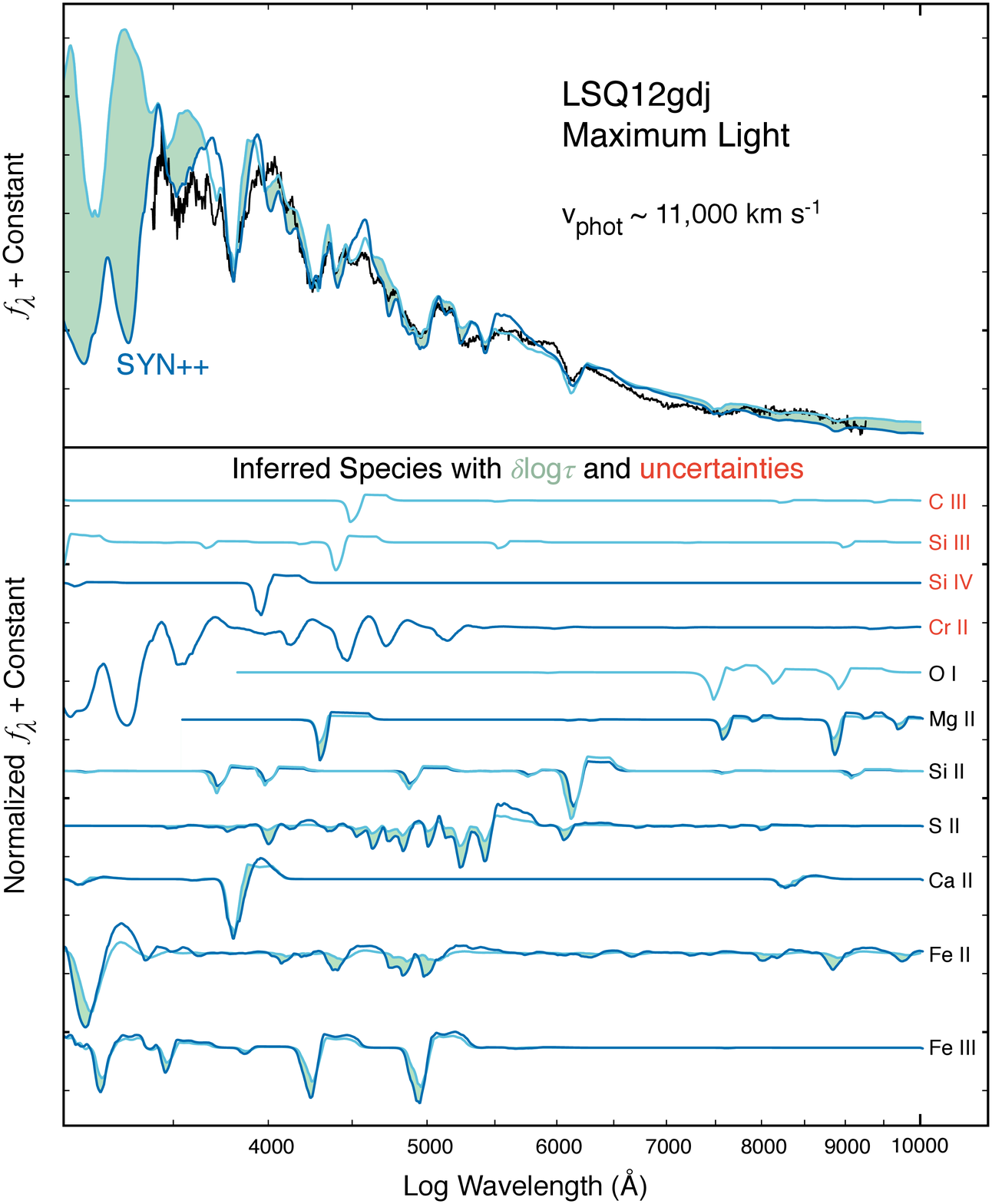}}
\caption{Maximum light \code{SYN++} fit comparisons between two best fits,
with and without the inclusion of \ion{Cr}{2}.  Species listed in red denote
degenerate solutions for some observed absorption features, while subsequent
uncertainties associated with respective line strengths ($\log\, \tau$)
between our two best fits are represented as a band of aquamarine.}
\label{fig:synow}
\end{center}
\end{figure*}

To identify various line features in a more comprehensive manner,
we fit the maximum-light spectrum of LSQ12gdj using \code{SYN++}
\citep{synapps}, shown in Figure~\ref{fig:synow}.  While LSQ12gdj displays
many features typical of SNe~Ia near maximum light, our fit also suggests
contributions from higher ionization species, e.g., \ion{C}{3} $\lambda$4649
over \ion{Fe}{2}/\ion{S}{2} absorption complexes, or \ion{Si}{3} near
3650~\AA\ and 4400~\AA\ in the pre-maximum spectra.  These identifications,
though tentative (labelled in red in Figure~\ref{fig:synow}), are consistent
with spectroscopic behaviors seen in shallow-silicon events prior to maximum
light \citep{branch06}.  The suggestion of \ion{C}{3}~$\lambda4649$ near
18,000~\kms\ is tantalizing, but ambiguous, and no corresponding
\ion{C}{2}~$\lambda 6580$ absorption is evident.
\ion{Cr}{2} is an intriguing possibility, since it provides
a better fit in the bluest part of the spectrum and simultaneously
contributes strong line blanketing in the unobserved UV part of the spectrum;
such line blanketing is in line with the sharp cutoff of our photometry-based
SED in the \emph{Swift} bands (see \S\ref{subsec:bolo}).  However, given the
degeneracies involved in identifying highly blended species, we do not
consider \ion{Cr}{2} to have been definitively detected in LSQ12gdj.

\begin{figure}
\begin{center}
\resizebox{0.5\textwidth}{!}{\includegraphics{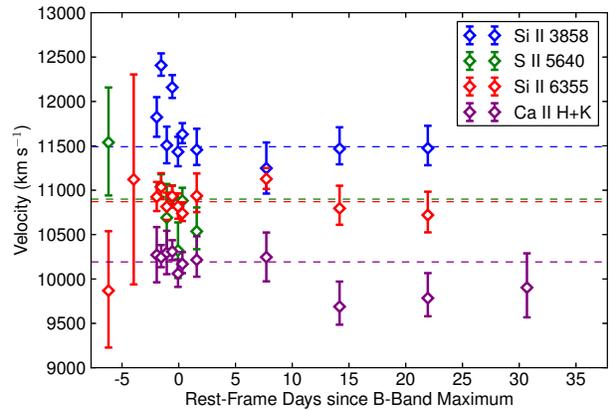}}
\caption{Blueshift velocities of absorption minima of intermediate-mass
element lines in the LSQ12gdj spectral time series.  Horizontal dashed lines
indicate the median velocity.  Asymmetric error bars represent the
68\% CL region for the absorption line minimum.}
\label{fig:vlines}
\end{center}
\end{figure}

We measure the absorption minimum velocities of common lines in a less
model-dependent way using a method similar to \cite{scalzo12}.  We resample
each spectrum to $\log(\lambda)$, i.e., to velocity space, then smooth it
with wide (``continuum''; $\sim 75000$~\kms)
and narrow (``lines''; $\sim 3500$~\kms) third-order Savitzsky-Golay filters,
which retain detail in the intrinsic line shapes more effectively than
rebinning or conventional Gaussian filtering.  After dividing out the
continuum to produce a smoothed spectrum with only line features, we measure
the absorption minima and estimate the statistical errors by Monte Carlo
sampling.  We track the full covariance matrix of the spectrum from the
original reduced data to the final smoothed version, and use its Cholesky
decomposition to produce Monte Carlo realizations.  We add in quadrature
a systematic error equal to the RMS spectrograph resolution, which may affect
the observed line minimum since we are not assuming a functional form
(e.g. a Gaussian) for the line profile.
The resulting velocities are shown in Figure~\ref{fig:vlines}.
In calculating velocities from wavelengths, we assume nearby component
multiplets are blended, with the rest wavelength of each line being the
$g$-weighted rest wavelengths of the multiplet components, although this
approximation may break down for some lines (see Figure~\ref{fig:linezoom}).

LSQ12gdj shows slow velocity evolution in the absorption minima of
intermediate mass elements, again characteristic of SN~1991T
\citep{phillips92} and other candidate super-Chandrasekhar-mass events
with 1991T-like spectra \citep{scalzo10,scalzo12}.
At early times, familiar absorption features of intermediate-mass elements
are either ambiguously identified or too weak for their properties to be
measured reliably, but come clearly into focus by maximum light.
Before maximum, the measured velocities for \ion{Si}{2}~$\lambda 3858$
differ by as much as 1000~\kms\ between neighboring WiFeS and CSP spectra.
The most likely source of the discrepancy is systematic error in the
continuum estimation for this shallow line near the blue edge of each
spectrograph's sensitivity, since the relative prominence of the local maxima
on either side of the line differ between CSP and WiFeS.  For other line
minima, measurements from CSP and from WiFeS are consistent with each other
within the errors.
For both \ion{Si}{2}~$\lambda 3858$ and \ion{Si}{2}~$\lambda 6355$,
$\vdot < 10$~\kms\ from maximum light until those lines become fully blended
with developing \ion{Fe}{2} lines more than three weeks past $B$-band maximum.
The \ion{Si}{2} plateau velocity is higher ($\sim 11000$~\kms) than any of the
\citet{scalzo12} SNe.
The velocity of \ion{Ca}{2}~H+K seems to decrease by about 500~\kms\
between day~$+7$ and day~$+14$, but on the whole it remains steady near
10000~\kms, with a velocity gradient consistent with that of \ion{Si}{2}.
The \ion{S}{2} $\lambda\lambda 5454, 5640$ ``W'' feature,
which often appears at lower velocities than \ion{Si}{2}, also appears around
11000~\kms\ until blending with developing \ion{Fe}{2} features erases it.

\revised{
Such velocity plateaus are predicted by models with density enhancements
in the outer ejecta, resulting from interaction with overlying material at
early times \citep{qhw07}.  These models include the DET2ENV2, DET2ENV4,
and DET2ENV6 ``tamped detonations'' of \citet{hk96}, hereafter collectively
called ``DET2ENVN'', and the ``pulsating delayed detonations'' by the same
authors.  In a tamped detonation, the ejecta interact with an external
envelope; in a pulsating delayed detonation, an initial pulsation fails to
unbind the white dwarf, and a shock is formed when the outer layers fall back
onto the inner layers before the final explosion.  The asymmetric models of
\citet{maeda10b,maeda11} also produce low velocity gradients, although in this
case the density enhancement occurs only along the line of sight.}

\revised{
Alternatively, the plateau may trace not the density but the composition of
the ejecta, i.e., may simply mark the outer edge of the iron-peak element
core of the ejecta.  The relatively normal SN~Ia 2012fr \citep{childress13},
for example, also featured extremely narrow (FWHM $< 3000$~\kms) absorption
features.}  SN~2012fr showed prominent high-velocity \ion{Si}{2} absorption
features, making it incompatible with a tamped detonation explosion scenario,
since any SN ejecta above the shock velocity would have been swept into
the reverse-shock shell.  No signs of high-velocity absorption features
from Ca, Si, or S are clearly evident in LSQ12gdj, although we might expect
such material to be difficult to detect in shallow-silicon events like
LSQ12gdj \citep{branch06}.


\subsection{Maximum-Light Behavior, Colors, and Extinction}
\label{subsec:phot}

The reddening due to Galactic dust extinction towards the host of LSQ12gdj
is $E(B-V)_\mathrm{MW} = 0.021$~mag \citep{sf11}.  LSQ12gdj was discovered
on the outskirts of a spiral galaxy viewed face-on, so we expect minimal
extinction by dust in the host galaxy.  The equivalent width of \ion{Na}{1}~D
absorption is $0.05 \pm 0.03$~\AA\ \citet{maguire13}, also consistent with
little to no host galaxy extinction.  A fit to the \citet{csp10} version of
the Lira relation \citep{phillips99} to the CSP $B$ and $V$ light curves
suggests \mbox{$E(B-V)_\mathrm{host} = 0.02 \pm 0.08$}~mag, consistent
with zero.

To obtain more precise quantitative constraints for use in later modeling,
we apply a multi-band light curve Bayesian analysis method to the CSP
light curve of LSQ12gdj, trained on normal SNe~Ia with a range of decline
rates \citep{burns14}.  This method provides joint constraints on
$E(B-V)_\mathrm{host}$ and the slope $R_{V,\mathrm{host}}$ of a
\citet{cardelli} dust law.  We find
$E(B-V)_\mathrm{host} = 0.013 \pm 0.005$~mag and $R_V = 1.66 \pm 1.66$
(a truncated Gaussian with $0 < R_V < 10$),
with covariance $C(E(B-V)_\mathrm{host}, R_{V,\mathrm{host}}) = -0.0039$~mag.
We adopt these values for our analysis.

As in \citet{scalzo14a}, we perform Gaussian process (GP) regression
on the light curve of each individual band, using the Python module
\code{sklearn} \citep{sklearn}, as a convenient form of
interpolation for missing data.  Gaussian process regression is a machine
learning technique which can be used to fit generic smooth curves to data
without assuming a particular functional form; we refer the reader to
\citet{rw06} for more details.  Neighboring points on the GP fit are
covariant; we use a squared-exponential covariance function
$k(t,t') = e^{-(t-t')^2/2\tau^2}$, with $0.5 < \tau < 2.0$.
When performing the fit, we include an extra term
$\sigma_i^2 \delta(t - t_i)$ describing the statistical noise on
the observations at times $t_i$ with errors $\sigma_i$; we neglect this
noise term when predicting values from the fit.

Figure \ref{fig:multi-lc} shows the light curve of LSQ12gdj in all available
bands, $S$-corrected to the appropriate standard system:  LSQ; Swift UVOT
$uvw1$, $uvm2$, $uvw2$, and $ubv$; Landolt $BV$; and SDSS $griz$.
Using the CSP bands, the SiFTO light curve fitter
\citep{sifto} gives a light curve stretch $s = 1.13 \pm 0.01$
and MJD of $B$-band maximum $56253.4$.
The SALT2.2 light curve fitter \citep{guy07,guy10} gives consistent results
($x_1 = 0.96 \pm 0.05$, $c = -0.048 \pm 0.026$), though with a slightly
later date for $B$-band maximum (MJD = 56253.8).
Using one-dimensional GP regression fits to each separate band yields
$B$-band maximum at MJD 56252.5 (2012 Nov 21.5, which we adopt henceforth),
colour at $B$-band maximum $(B-V)_\mathrm{max} = -0.019 \pm 0.005$,
$\Delta m_{15}(B) = 0.74 \pm 0.01$, and peak magnitudes
$m_B = 15.972 \pm 0.004$, $m_V = 15.947 \pm 0.004$.
These errors are statisical only; systematic errors are probably around 2\%.
After correction for the mean expected reddening,
we derive peak absolute magnitudes $M_B = -19.78$, $M_V = -19.77$,
using a distance modulus $\mu = 35.60 \pm 0.07$ derived from the redshift
assuming a $\Lambda$CDM cosmology with $H_0 = 67.3$~\kms~Mpc$^{-1}$
\citep{planck13} and a random peculiar velocity of 300~\kms.

We find a fairly substantial ($\sim 0.2$~mag) mismatch between
\emph{Swift} $b$ and CSP $B$, and between \emph{Swift} $v$ and CSP $V$,
near maximum light; at later times, \emph{Swift} and CSP observations agree
within the errors (of the \emph{Swift} points).  The shape of the light curve
is strongly constrained by CSP data, so we use CSP data in constructing the
bolometric light curve at a given phase when both \emph{Swift} and CSP
observations are available.

The second maximum in the CSP $i$ light curve appears earlier ($+25$~days)
than expected for LSQ12gdj's $\Delta m_{15}(B)$ \citep[$+30$~days][]{csp10}.
The contrast of the second maximum is also fairly low, with a difference of
$-0.63$~mag with the first maximum and $+0.20$~mag with the preceding minimum.
Similar behavior is seen in LCOGT $z$.  These properties are typical of
low-\nickel\ explosions among the Chandrasekhar-mass models of
\citet{kasen06}, difficult to reconcile with LSQ12gdj's high luminosity.
The low contrast persists even when CSP $i$ and LCOGT $z$ are $S$-corrected
to Landolt $I$ for more direct comparison with \citet{kasen06}.
If LSQ12gdj synthesized a high mass of \nickel, comparison with the models
of \citet{kasen06} suggests that LSQ12gdj has substantial mixing of \nickel\
into its outer layers (as we might expect from its spectrum), a high yield
of stable iron-peak elements, or both.

Fitting a $t^{2.0 \pm 0.2}$ rise to LSQ points more than a week before
$B$-band maximum light suggests an explosion date of MJD 56235.9, giving a
$B$-band rise time of $16.2 \pm 0.3$~days.
The pre-explosion upper limits are compatible with a $t^2$ rise,
but do not permit LSQ12gdj to be visible much before $B$-band phase $-16$.
\revised{
The $t^2$ functional form is at best approximate; depending on how \nickel\
is distributed in the outer layers, the finite diffusion time of photons
from \nickel\ decay could result in a ``dark phase'' before the onset of
normal emission \citep{hachinger13,pn13,pn14,mazzali14}.}  Nevertheless,
LSQ12gdj has a somewhat shorter visible rise than other SNe~Ia with
similar decline rates \citep{ganesh11}, and much shorter than the 24-day
visible rise of SN~2007if \citep{scalzo10} determined by the same method.


\subsection{Bolometric Light Curve}
\label{subsec:bolo}

We construct a bolometric light curve for LSQ12gdj in the rest-frame
wavelength range 1550--23100~\AA\ using the available photometry, as follows.

We first generate quasi-simultaneous measurements of all bandpasses at
each epoch in Tables~\ref{tbl:photometry-opt} and \ref{tbl:photometry-swift}.
We interpolate the values of missing measurements at each using the GP fits
shown in Figure~\ref{fig:multi-lc}.  After observations from a given band
cease because the SN is no longer detected against the background,
we estimate upper limits on the flux by assuming that the mean colours of
the SN do not change since the last available observation --- in particular
that the SN does not become bluer in the \emph{Swift} bands.  If the last
measurement in band $j$ was taken at time $t_{\mathrm{last},j}$,
then at all future times $t_i$ we form the predictions
\begin{equation}
m_{i,j,j'} = m_{i,j'} + (m_{\mathrm{last},j} - m_{\mathrm{last,j'}})
\end{equation}
and set the upper limit $m_{i,j}$ by averaging $m_{i,j,j'}$ over all
remaining bands $j'$.  We estimate, and propagate, a systematic error
on this procedure by taking the standard deviation of $m_{i,j,j'}$ over
all remaining bands $j'$.  All other bands used for this construction
(Swift $b$, CSP $BVri$ and LCOGT $z$) have adequate late-time coverage.
The projected values are consistent with upper limits measured from
non-detection in those bands, and the contribution of these bands
to the bolometric luminosity is small ($< 5\%$) at late times.

At each epoch, we construct a broad-band SED of LSQ12gdj in the observer
frame using the natural-system transmission curves, and then de-redshift it
to the rest frame, rather than computing full $K+S$-corrections for all of
our broadband photometry.  Since we have no detailed UV or NIR time-varying
spectroscopic templates for LSQ12gdj, full $K+S$-corrections are not feasible
for all of the \emph{Swift} bands; since we need only the overall
bolometric flux over a wide wavelength range instead of rest-frame
photometry of individual bands, they are not strictly necessary.
We have no NIR photometry of LSQ12gdj either, so we use the NIR template
described in \citet{scalzo14a} to predict the expected rest-frame
magnitudes in $YJHK$ band for a SN~Ia with $x_1 = 1$.  The size of the
correction ranges from a minimum of 7\% near maximum light to 27\% around
35~days after maximum light, comparable to the observed NIR fractions for
SN~2007if \citep{scalzo10} and SN~2009dc \citep{taub13}.

\renewcommand{\vec}[1]{\textbf{\textit {#1}}}

To determine a piecewise linear observer-frame broadband SED at each epoch,
$F(\lambda_j)$, evaluated at the central wavelength $\lambda_j$
of each band $j$, we solve the linear system
\begin{equation}
\frac{\int F_n(\lambda) T_j(\lambda) \, d\lambda}
     {\int T_j(\lambda) \, d\lambda} = 10^{-0.4(m_j)},
\label{eqn:sedflux}
\end{equation}
where $m_j$ is the observed magnitude, and $T_j(\lambda)$ the filter
transmission, in band $j$.  The system is represented as a matrix equation
$\mathbf{A} \vec{x} = \vec{b}$, where \vec{x} and \vec{b} give the flux
densities and observations, and $\mathbf{A}$ is the matrix of a linear
operator corresponding to the process of synthesizing photometry.
We discretize the integrals via
linear interpolation (i.e., the trapezoid rule) between the wavelengths at
which the filter transmission curves are measured.  We solve the system using
nonlinear least squares to ensure positive fluxes everywhere.  The Swift
$uvw1$ and $uvw2$ bands have substantial red leaks \citep{breeveld11},
but the red-leak flux is strongly constrained by the optical observations,
and we find our method can reproduce the original \emph{Swift} magnitudes
to within the errors.  We exclude Swift $B$ and $V$ when higher-precision
CSP $B$ and $V$ measurements are available, covering similar wavelength
regions.  To convert this observer-frame SED to the rest frame, we follow
\citet{nkp02}:
\begin{equation}
f_\lambda^z(\lambda) \, d\lambda =
   \frac{d\lambda}{1+z} \, f_\lambda \left( \frac{\lambda}{1+z} \right).
\end{equation}
We integrate the final SED in the window 1550--23000~\AA\ to obtain the
bolometric flux.  Simulating this procedure end-to-end using $UBVRI$
synthetic photometry on SN~Ia spectra from the BSNIP sample \citep{bsnip}
with phases between $-9$ and $+460$~days, we find that (for zero reddening)
we can reproduce the 3250--8000~\AA\ quasi-bolometric flux to within 1\%
(RMS).  We add this error floor as a systematic error in quadrature to each
of our bolometric flux points.

To account for Milky Way and host galaxy extinction, we make bolometric light
curves for a grid of possible $E(B-V)$ and $R_V$ values, fixing $R_V = 3.1$
for the Milky Way contribution.  We sample $E(B-V)_\mathrm{host}$ in 0.01~mag
steps from 0.00--0.20~mag, and we sample $R_{V,\mathrm{host}}$
in 0.2 mag/mag steps from 0.0--10.0.  We interpolate the light curves
linearly on this grid as part of the Monte Carlo sampling described in
\S\ref{subsec:massrecon}, applying our prior on $E(B-V)_\mathrm{host}$ and
$R_{V,\mathrm{host}}$ given in \ref{subsec:phot} and constraining their
values to remain within the grid during the sampling.

\begin{figure}
\resizebox{0.5\textwidth}{!}{\includegraphics{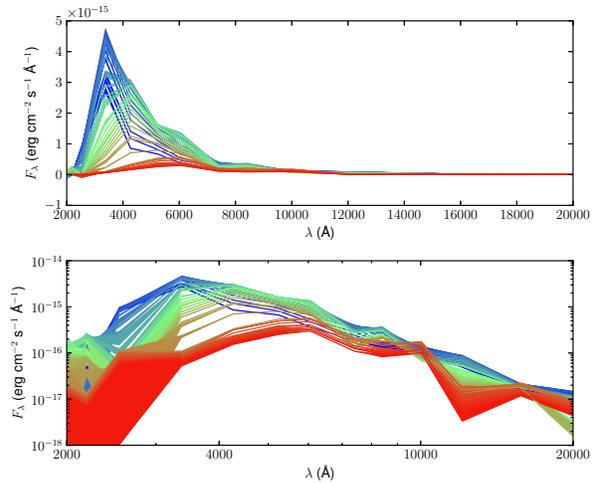}}
\caption{Rest-frame low-resolution SED time series of LSQ12gdj,
1550--23100~\AA, constructed from \emph{Swift} UV + CSP $BVr'i'$ photometry.
SED colours show the phase, ranging from blue ($-10$~days) to red
($+45$~days).  Top: linear scale; bottom: log scale.  Colored bands represent
the 1$\sigma$ confidence region around the mean.}
\label{fig:sedflux}
\end{figure}

Figure \ref{fig:sedflux} shows the resulting time-dependent SED of LSQ12gdj
for zero host galaxy reddening.  The peak wavelength changes steadily as
the ejecta expand and cool, making \emph{Swift} $u$ the most luminous
band at early phases.  Although a significant fraction of the flux is emitted
bluewards of 3300~\AA, the flux density cuts off sharply bluewards of
\emph{Swift} $uvm2$. Less than 1\% of the flux is emitted bluewards of
2300~\AA\ at all epochs, and our SED in these regions is consistent with
statistical noise.  This behavior is inconsistent with simply being the
Rayleigh-Jeans tail of a hot blackbody.  Although we have no UV spectroscopy
of LSQ12gdj, we expect the sharp cutoff blueward of 3000~\AA\ for the entire
rise of the SN to be formed by line blanketing from iron-peak elements
(e.g. \ion{Cr}{2}, as in Figure~\ref{fig:synow}),
as is common in SNe~Ia.

\begin{figure}
\resizebox{0.5\textwidth}{!}{\includegraphics{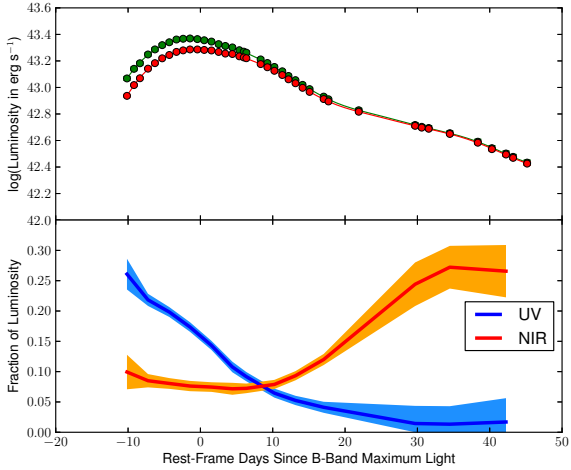}}
\caption{Rest-frame bolometric light curve of LSQ12gdj obtained by
integrating the SED of Figure~\ref{fig:sedflux}.
Top:  Bolometric light curve 3300--23100~\AA\ (red) and 1550--23100~\AA\
(green) representing the results with and without Swift UV, including
Gaussian process regression fit (connecting curves).
Bottom:  Fraction of bolometric flux bluewards of 3300~\AA\ (``UV'')
and redwards of 8800~\AA\ (``NIR'', estimated from a template);
the solid curves show the mean behavior, and the lighter bands show
the 1$\sigma$ confidence region around the mean.}
\label{fig:bolo-lc}
\end{figure}

Figure \ref{fig:bolo-lc} shows the bolometric light curve, together with
Gaussian process regression fits.  Like other candidate
super-Chandrasekhar-mass SNe~Ia observed with \emph{Swift} \citep{brown14},
LSQ12gdj is bright at UV wavelengths from the earliest phases.
Up to 27\% of the bolometric flux is emitted blueward of 3300~\AA\
at day~$-10$, decreasing to 17\% at $B$-band maximum light
and to $< 5\%$ by day~$+20$.  After day~$+20$, the SN is no longer detected
in the \emph{Swift} bands, so the small constant fraction reflects our
method of accounting for missing data (with large error bars).
For comparison, in the well-observed normal SN~Ia~2011fe \citep{pereira13},
at most 13\% of the luminosity is emitted blueward of 3400~\AA, reaching
this point 5~days before $B$-band maximum light; the UV fraction is 9\%
at day~$-10$, and only 3\% at day~$-15$.  UV flux contributes only 2\% of
SN~2011fe's total luminosity by day +20, and continues to decline thereafter.

\begin{table*}
\center
\caption{Ground-based photometry of LSQ12gdj in the natural systems of the
         Swope and LCOGT telescopes}
\newcommand{\ctr}[1]{\multicolumn{1}{c}{#1}}
\newcommand{\ergps}{erg~s$^{-1}$}
\begin{tabular}{rrrrrr}
\hline 
\ctr{Phase$^a$} & \ctr{$L_\mathrm{bol}^b$} & \ctr{$\sigma_{L,\mathrm{stat}}^c$} & \ctr{$\sigma_{L,\mathrm{sys}}^d$}
                & \ctr{$\sigma_{L,\mathrm{tot}}^e$} & \ctr{$f_\mathrm{NIR}^f$} \\
\ctr{(days)} & \ctr{($10^{43}$ \ergps)} & \ctr{($10^{43}$ \ergps)} & \ctr{($10^{43}$ \ergps)} & \ctr{($10^{43}$ \ergps)} & \\
\hline 
$-10.2$ & 1.252 & 0.033 & 0.025 & 0.041 & 0.094 \\
$ -9.2$ & 1.472 & 0.026 & 0.029 & 0.039 & 0.088 \\
$ -8.4$ & 1.616 & 0.030 & 0.032 & 0.044 & 0.087 \\
$ -7.2$ & 1.882 & 0.018 & 0.037 & 0.041 & 0.081 \\
$ -6.3$ & 2.051 & 0.017 & 0.041 & 0.044 & 0.080 \\
$ -5.3$ & 2.208 & 0.017 & 0.044 & 0.047 & 0.078 \\
$ -4.3$ & 2.320 & 0.018 & 0.046 & 0.049 & 0.077 \\
$ -3.3$ & 2.437 & 0.035 & 0.048 & 0.060 & 0.074 \\
\hline 
\end{tabular}
\medskip \\
\flushleft
The full version of this table can be found online. \\
$^a$ Phase given in rest-frame days since $B$-band maximum light.\\
$^b$ Luminosity estimate assumes Milky Way galaxy dust extinction from the dust maps of \citet{sf11},
     baseline host galaxy dust extinction ($R_{V,\mathrm{host}} = 1.66$, $E(B-V)_\mathrm{host} = 0.013$),
     and a distance modulus mu = 35.60 derived from the Planck LCDM cosmology \citep{planck13}. \\
$^c$ Statistical error includes measurement errors on imaging photometry. \\
$^d$ Systematic error includes measured discrepancies between imaging photometry and synthetic photometry
     of the trapezoidal SED, with a 2\% floor based on Monte Carlo simulations of how well we can recover
     the true bolometric flux of a full SN spectrum.  Uncertainties in the NIR correction template
     \citep{scalzo14a} are not considered here, but do not exceed 2\% during the epochs used in our
     light curve fitting. \\
$^e$ Quadrature sum of statistical and systematic errors. \\
$^f$ Estimated fraction of bolometric luminosity in near-infrared wavelengths. \\

\label{tbl:bolophot}
\end{table*}


\section{PROGENITOR PROPERTIES}
\label{sec:discussion}

In this section we perform some additional analysis \revised{to constrain
properties of the LSQ12gdj progenitor:  the ejected mass, the \nickel\ mass,
and the physical configuration of the CSM envelope (if one is present).}
We fit the bolometric light curve
in \S\ref{subsec:massrecon} to infer the ejected mass and place rough
constraints on trapped radiation from interaction with a compact envelope.
In \S\ref{subsec:chevalier} we attempt to constrain the impact of interaction
with an extended CSM wind, including constraints on CSM mass based on
blueshifted \ion{Na}{1}~D absorption \citep{maguire13} and a light curve
comparison to known CSM-interacting SNe~Ia.  Finally, in \S\ref{subsec:2007if}
we consider the implications of our findings for the more established
super-Chandrasekhar-mass SNe~Ia, including SN~2007if and SN~2009dc.


\subsection{Ejected Mass, \nickel\ Mass, and Trapped Thermal Energy
            from Interaction with a Compact CSM}
\label{subsec:massrecon}

LSQ12gdj has excellent UV/optical coverage from well before maximum to over
40~days after maximum, allowing us to model it in more detail than possible
for many other SNe~Ia.  We use the \code{bolomass} code
(Scalzo \etal, in prep), based on a method applied to
other candidate super-Chandrasekhar-mass SNe~Ia \citep{scalzo10,scalzo12},
as well as normal SNe~Ia \citep{scalzo14a}.  The method constrains the
\nickel\ mass, \MNi, and the ejected mass, \MWD, using data both near maximum
and at late times, when the SN is entering the early nebular phase.

\code{bolomass} uses the \citet{arnett82} light curve model, including as
parameters \revised{\MNi\ and} the expected time $t_0$ at which the ejecta
become optically thin to \cobalt\ gamma rays.  However, \code{bolomass} also
calculates the expected transparency of the ejecta to gamma rays from
\cobalt\ decay at late times, using the formalism of \citet{jeffery99}
together with a 1-D parametrized model \{$\,\rho(v)$,~$\mathbf{X}(v)\,$\}
of the density and composition structure as a function of the ejecta
velocity $v$.  The effective opacity for Compton scattering (and subsequent
down-conversion) of \cobalt-decay gamma rays in the optically thin limit
\citep{swartz95} is much more precisely known than optical-wavelength line
opacities near maximum light \citep{kmh93}; this allows \code{bolomass} to
deliver robust, quantitative predictions, \revised{avoiding uncertainties
associated with scaling arguments or assumptions about the optical-wavelength
opacity.} \code{bolomass} uses the affine-invariant
Monte Carlo Markov Chain sampler \code{emcee}
\citep{emcee} to sample the model parameters and marginalize over nuisance
parameters associated with systematic errors, subject to a suite of priors
which encode physics from contemporary explosion models.

The \citet{arnett82} light curve model includes as parameters not only
\revised{\MNi\ and $t_0$, but the effects of photon diffusion on the overall
light curve shape,} the initial thermal energy $E_\mathrm{th}$ of the ejecta
and the finite size $R_0$ of the exploding progenitor.  They enter through
the dimensionless combinations
\begin{eqnarray}
y \!\! & = & \!\! \frac{\trise}{2 \tau_\mathrm{Ni}}
         = \frac{\trise}{17.6~\mathrm{days}}, \\
w \!\! & = & \!\! \frac{2R_0}{\trise\vKE}
         \sim \frac{R_0}{10^{15}~\mathrm{cm}}.
\end{eqnarray}
While $w \ll 1$ for white dwarfs, allowing $w$ to float in this case may
help us estimate the contribution of trapped radiation from interaction
with a compact, hydrogen-free CSM envelope which might otherwise be invisible;
this formalism is not appropriate for an ongoing shock interaction.

\begin{figure}
\begin{center}
\resizebox{0.5\textwidth}{!}{\includegraphics{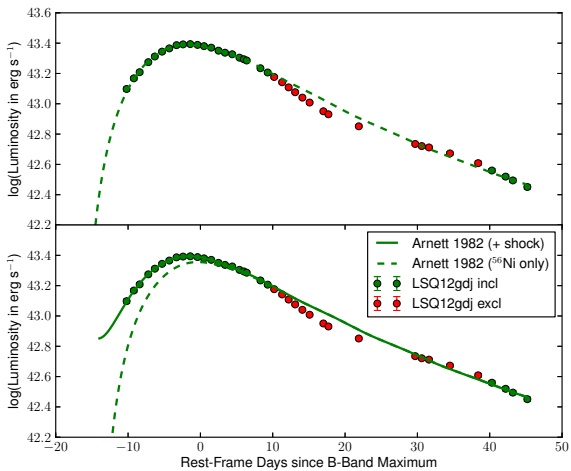}}
\caption{Fits of the \citet{arnett82} light curve model to the bolometric
light curve of LSQ12gdj, including the trapping of thermalized kinetic energy
from an interaction with a compact carbon-oxygen envelope, in the context of
the tamped detonation scenario.  Top:  Zero initial radius.
Bottom:  Initial radius and thermal energy determined by the fit, showing the
full fit (solid curve) and the contribution due only to radioactive decay
(dashed curve).  Red symbols mark points in the transition regime from
photospheric to early nebular phase, which have been excluded from the fit.}
\label{fig:arnett-fits}
\end{center}
\end{figure}

\citet{arnett82} points out that \revised{the assumptions of constant opacity
and a sharp photosphere in the expanding SN ejecta} break down
between maximum light and late phases, when the SN is in transition from full
deposition of radioactive decay energy to the optically thin regime.
We therefore exclude light curve points between 10~days and 40~days after
$B$-band maximum light, and find that the \citet{arnett82} light curve model
provides an excellent fit \revised{($< 1\%$ dispersion)}
to the remaining points.

\revised{The Arnett model also does not treat the \nickel\ distribution
in detail, assuming only that it is centrally concentrated, which prevents it
from accurately predicting the light curve shape at very early times
\citep{hachinger13,pn13,pn14,mazzali14}, as mentioned in \S\ref{subsec:phot}.
For example, if LSQ12gdj's actual rise time is longer
than the predictions of our model, the actual \nickel\ mass could be larger;
a 2-day ``dark phase'' could increase the inferred \nickel\ mass by
up to 25\% relative to our estimate.  \citet{pe00} consider the impact of
large amounts of \nickel\ at high velocities on the light curve:  for a
uniform \nickel\ distribution, the lower Compton depth results in a shorter
rise time (by as much as 3 days), but also lower $\alpha$ (0.85) as some of
the radioactive energy simply leaks out of the ejecta, and these effects
roughly balance for our \nickel\ estimates.
More detailed modeling of our spectroscopic time series may
provide further information about the true distribution of \nickel.}

Figure \ref{fig:arnett-fits} shows \revised{two possible fits to}
the bolometric light curve of LSQ12gdj.  When we fix $w = 0$
and consider only \revised{luminosity from radioactivity,} we recover
$\trise = 16.4$~days, in agreement with the $t^2$ fit to the early-phase
LSQ data, and $\MNi = 1.00$~\Msol.
Allowing $w$ to float reveals a second possible solution,
in which trapped thermal energy contributes
around 10\% of the luminosity at maximum light.  The fit has $w = 0.013$
and $E_\mathrm{th} = 6 \times 10^{50}$~erg, and has a significantly
shorter rise time $\trise = 14.1$~days, exploding just before the
initial detection by LSQ.  This value of $w$ corresponds to an effective
radius of roughly $10^{13}$~cm, more extended than the envelopes in the
DET2ENVN series \citep{kmh93} \revised{but comparable to those predicted by
\citet{shen12}.}  The amount of thermalized kinetic energy is compatible
with the formation of a reverse-shock shell near 10000~\kms in a tamped
detonation \revised{or pulsating delayed detonation.}
Importantly, the trapped radiation contributes most at early times and
around maximum light, but disappears on a light curve width timescale,
just as suggested by \citet{hachinger12} and \citet{taub13} in the case of
SN~2009dc.  The late-time behavior is the same as for the radioactive-only
case, and the best-fit \nickel\ mass is 0.88~\Msol.  The reduced chi-squares
for both fits are very low (0.47 for $w = 0$ versus 0.15 for $w > 0$),
so that while the $w > 0$ fit is technically favored, both are consistent
with our observations.

The ejected mass estimate depends on the actual density structure and
\nickel\ distribution in the ejecta.  We consider two possible functional
forms for the density structure, one which depends exponentially on
velocity and one with a power-law dependence, as in \citet{scalzo14a}.
We parametrize the stratification of the ejecta by a mixing scale \aNi\
\citep{kasen06}, and consider stratified cases with $\aNi = 0.1$
and well-mixed cases with $\aNi = 0.5$.
The detailed model-dependence of the trapping of radiation near maximum
light is often factored out into a dimensionless ratio $\alpha$
\citep{nugent95,howell06,howell09} of order unity, by which the rough
``Arnett's rule'' estimate of \MNi\ from the maximum-light luminosity is
divided.  Here we use the Arnett light curve fit directly to estimate
the \nickel\ mass and the amount of thermalized kinetic energy trapped
and released, resulting in an effective $\alpha$ between 1.0 and 1.1 for
LSQ12gdj.  For the $w = 0$ Arnett formalism, $\alpha = 1$ by construction,
ignoring both opacity variation in the ejecta and/or less than complete
gamma-ray deposition near maximum light \citep{blondin13}.
\revised{In one-dimensional, Chandrasekhar-mass delayed detonation models
\citep[e.g.][]{kmh93,hk96}, a high central density may enhance neutronization
near the center of the ejecta, creating a \nickel-free ``hole''.  Some
evidence for such a hole is found in late-time spectra of SNe~Ia
\citep{hoflich04,motohara06,mazzali07}, and in some multi-dimensional
simulations \citep{maeda10a}, while other simulations do not support such
an effect \citep{krueger12,seitenzahl13}.  We consider cases both with and
without \nickel\ holes due to neutronization, as in \citet{scalzo14a}.}
In all reconstructions, we allow the unburned carbon/oxygen fraction,
the envelope size, and the thermalized kinetic energy to float freely
\revised{to reproduce the observed bolometric light curve, including
variations in the rise time produced by changes in these parameters.}

\begin{table}
\center
\caption{Ejected mass and \nickel\ mass of LSQ12gdj under various priors}
\begin{tabular}{lccrrrr}
\hline 
   Run &
   $\rho(v)$\nb{a} &
   $a_\mathrm{Ni}$\nb{b} &
   $\vKE$\nb{c} (\kms) &
   $\MWD/\Msol$\nb{d} &
   $P_\mathrm{SCh}$\nb{e} \\
\hline 
A & pow3x3 & 0.5 & $10390^{+634}_{-284}$ & $1.56^{+0.13}_{-0.08}$ & 0.992 \\[0.3ex]
B & pow3x3 & 0.1 & $10484^{+613}_{-295}$ & $1.51^{+0.10}_{-0.07}$ & 0.953 \\[0.3ex]
C & exp    & 0.5 & $10713^{+503}_{-250}$ & $1.43^{+0.08}_{-0.05}$ & 0.691 \\[0.3ex]
D & exp    & 0.1 & $10939^{+485}_{-292}$ & $1.38^{+0.06}_{-0.07}$ & 0.354 \\[0.3ex]
\hline 
\end{tabular}
\medskip \\
\flushleft
\nb{a}~{Assumed density profile as a function of ejecta velocity: \\
        ``exp'' $\propto \exp(-\sqrt{12}v/\vKE)$, as in 1-D explosion models. \\
        ``pow3x3'' $\propto [1 + (v/\vKE)^{3}]^{-3}$, similar to 3-D models cited in \citet{scalzo14a}.} \\
\nb{b}~{Assumed width of the mixing layer near the iron-peak core boundary,
        in mass fraction; 0.1 is highly stratified while 0.5 is well-mixed \citep{kasen06}.} \\
\nb{c}~{Kinetic energy scaling velocity.}
\nb{d}~{Total ejected mass.} \\
\nb{e}~{Fraction of the integrated probability density lying above $\MWD = 1.4$~\Msol.}

\label{tbl:massrecon}
\end{table}

Table~\ref{tbl:massrecon} shows the inferred \MWD\ and probability of
exceeding the Chandrasekhar limit for four different combinations of
priors, marginalizing over the full allowed range of $w$ and $E_\mathrm{th}$.
The full Monte Carlo analysis
robustly predicts $\trise = 16 \pm 1$~days and $\MNi = 0.96 \pm 0.07$~\Msol.
The thermalized kinetic energy is constrained to be less than about
$10^{51}$ erg; this maximum value results in ejecta with a \emph{maximum}
velocity around 10000~\kms, roughly consistent with our observations.
\revised{Models which allow \nickel\ holes shift \nickel-rich material to
lower Compton optical depths, requiring more massive ejecta to reproduce the
late-time bolometric light curves; for LSQ12gdj, this effect is small since
the favored super-Chandrasekhar-mass solutions are rapidly rotating
configurations with low central density \citep{yl05}.  If LSQ12gdj in fact
had a high central density with corresponding \nickel\ hole,
its ejected mass should be at least 0.1~\Msol\ higher than what we infer.
Since \nickel\ makes up such a large fraction of the ejecta in any case,
there is little difference between the well-mixed models and the stratified
models.} Models with power-law density profiles have \MWD\ larger by about
0.14~\Msol\ than models with exponential density profiles;
\revised{this is the largest predicted uncertainty in our modeling.}

\revised{The uncertainty from the unknown ejecta density profile is not
easily resolved.  Although \code{bolomass} can model any user-defined
spherically symmetric density structure, the light curve is sensitive
primarily to the total Compton scattering optical depth, and not directly
to the actual ejecta density profile, except for the most highly disturbed
density structures.  \citet{scalzo14a} showed that assuming an exponential
density profile led to biases in the reconstructed mass for multi-dimensional
explosion models best represented by power laws.  Our judgment as to whether
LSQ12gdj is actually super-Chandrasekhar-mass thus hinges mostly on which
density profile we believe to be more realistic.}

If LSQ12gdj is indeed a tamped detonation, it is probably (slightly)
super-Chandrasekhar-mass, and could be explained by a Chandrasekhar-mass
detonation inside a compact envelope of mass around 0.1~\Msol.
If all of LSQ12gdj's luminosity is due to radioactive energy release,
it could be (slightly) sub-Chandrasekhar-mass, a good candidate for a
double detonation \citep{ww94,fink10} of about 1.3~\Msol, a conventional
Chandrasekhar-mass near-pure detonation \citep{blondin13,seitenzahl13},
\revised{or a pulsating delayed detonation \citep{kmh93,hk96}.}

Interestingly, the DDC0 delayed-detonation model of \citet{blondin13}
has a rise time of 15.7~days, very close to the value we observe.
\revised{The 1.4-\Msol\ tamped detonation 1p0\_0.4 of \citet{raskin14} also
comes close to our expected scenario, with a roughly spherical helium envelope
that has been thermalized in the merger interaction.  The spectra near maximum
are blue with shallow features.  The envelope is compact,
with a density profile following a $r^{-4}$ power law, but could expand to
$\sim 10^{13}$ cm if the detonation of the white dwarf primary is delayed
after the merger event \citep{shen12}.}


\subsection{Constraints on Ongoing Shock Interaction with an Extended CSM}
\label{subsec:chevalier}

We also address the question of whether LSQ12gdj might be undergoing shock
interaction with a hydrogen-poor extended wind, adding luminosity to its
late-time light curve.  The ``Ia-CSM'' events, such as SN~2002ic
\citep{hamuy03}, SN~2005gj \citep{snf2005gj,prieto07}, SN~2008J
\citep{taddia12}, and PTF11kx \citep{dilday12}, have spectra which seem
to be well-fit by a combination of a 1991T-like SN~Ia spectrum, a broad
continuum formed at the shock front, and narrow H$\alpha$ lines formed
in photoionized CSM \citep{silverman13a,silverman13b,leloudas13}.
A hydrogen-poor extended CSM could produce
pseudocontinuum luminosity and and weaken absorption lines via toplighting
\citep{branch00}, while not producing any distinctive line features itself,
although a very massive envelope could in principle produce carbon or
oxygen lines \citep{benami14}.

\begin{figure}
\begin{center}
\resizebox{0.5\textwidth}{!}{\includegraphics{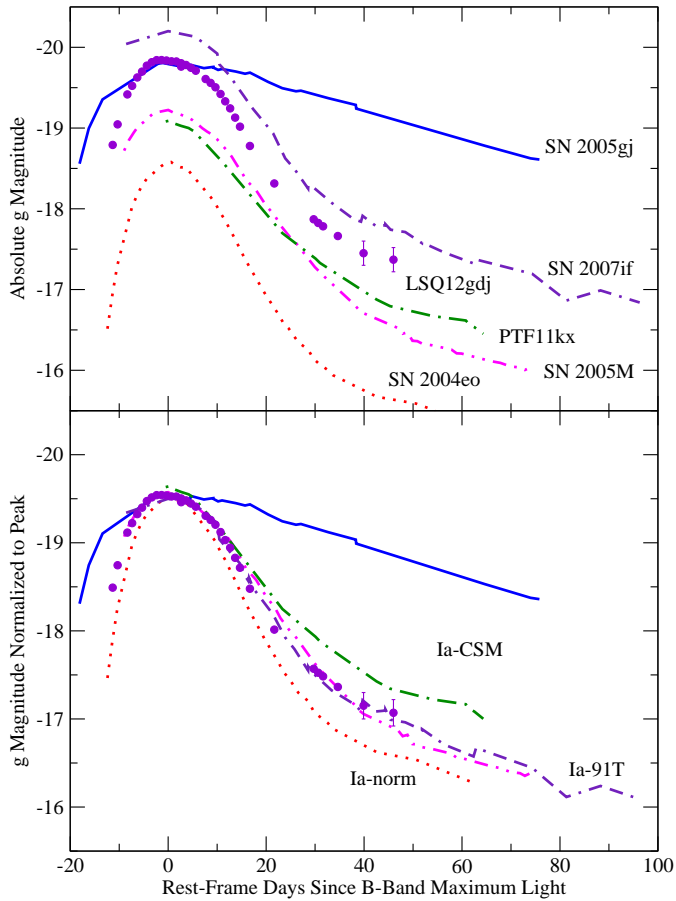}}
\caption{Comparison of $g$-band light curves of peculiar SNe~Ia.
The CSP and \emph{Swift} $B$-band light curves of LSQ12gdj (filled circles)
have been $S$-corrected to $g$ band using the spectra in this paper.
References for the other light curves include:
SN~2005gj, from \citet{prieto07} (solid curve);
PTF11kx, from \citet{dilday12} (dash-dot curve);
SN~2004eo and SN~2005M, from \citet{contreras10}
   (dotted and dash-dot-dot curves);
and SN~2007if, from \citet{cspdr2} and \citet{scalzo10}
   (dash-dash-dot curve).}
\label{fig:gmag-compare}
\end{center}
\end{figure}

Figure \ref{fig:gmag-compare} shows the $g$-band light curve of LSQ12gdj
alongside those of the Ia-CSM SN~2005gj and PTF11kx,
the super-Chandrasekhar-mass SN~2007if, the 1991T-like SN~2005M,
and the fast-declining, spectroscopically normal SN~2004eo. 
We choose $g$ for the comparison because it is the bluest band observed
(or synthesizable) in common for all of the SNe.
SN~2005gj, the clearest example of ongoing shock interaction,
declines extremely slowly, with far more luminosity at day~$+40$ and later
than any of the other SNe.  PTF11kx, a case of an intermediate-strength
shock interaction, has peak brightness comparable to the 1991T-like SN~2005M,
but shows a long tail of shock interaction luminosity and is up to 0.5~mag
more luminous than SN~2005M at day~$+60$.  By a year after explosion,
the spectrum of PTF11kx, like that of SN~2005gj, is dominated by
shock interaction signatures such as H$\alpha$ \citep{silverman13a},
rather than by \ion{Fe}{2} as for SN~2007if \citep{taub13}.

Despite having peak magnitudes that differ over a range of 1~mag,
SN~2005M, SN~2007if, and LSQ12gdj all have very similar post-maximum light
curve shapes, more consistent with each other than with the Ia-CSM.
This puts strong constraints on the density and geometry of any CSM present;
existing examples of Ia-CSM show that the light curve shapes can vary
dramatically according to the density and geometry of the surrounding medium.
In particular, none of the SNe~Ia-91T show an extended power-law tail to the
light curve, nor do they match expectations from radiation hydrodynamics
simulations of heavily enshrouded SNe~Ia in hydrogen-poor envelopes
\citep{fryer10,bs10}.

\newcommand{\Menv}{\ensuremath{M_\mathrm{env}}}
\newcommand{\vsh}{\ensuremath{v_\mathrm{sh}}}

We can derive a more quantitative upper limit on the presence of extended CSM
by searching for circumstellar \ion{Na}{1}~D absorption.
\citet{maguire13} observed LSQ12gdj with the XSHOOTER spectrograph on the
ESO Very Large Telescope at Paranal, finding narrow \ion{Na}{1}~D and
\ion{Ca}{2}~H+K absorption blueshifted at $-220$~\kms\ relative to the
recessional velocity of the LSQ12gdj host.
LSQ12gdj is one of a larger sample of SNe~Ia with blueshifted absorption
features studied in \citet{maguire13}.  While blueshifted \ion{Na}{1}~D
absorption would be expected statistically for a population of progenitors
surrounded by a CSM wind, of either single-degenerate \citep{sternberg11}
or double-degenerate \citep{shen12,rk13} origin, we have no way of knowing
whether such absorption is circumstellar for any individual SN~Ia, or
whether it arises from interstellarmaterial in the host galaxy.

We can nevertheless derive a conservative upper limit assuming all of the
absorption arises from hydrogen-rich CSM.  Since $\ewna$ near the host galaxy
redshift is small, there should be little CSM present around LSQ12gdj.
We use the \code{vpfit}\footnote{
VPFIT was developed by R. F. Carswell and can be downloaded for free at 
\url{http://www.ast.cam.ac.uk/~rfc/vpfit.html}.}
code to place an upper limit on the column density of \ion{Na}{1}
using the XSHOOTER spectrum from \citet{maguire13}, in the case in which
all \ion{Na}{1}~D absorption is circumstellar; we obtain
$N(\mathrm{\ion{Na}{1}}) < 4 \times 10^{11}~\mathrm{cm}^{-2}$.
For a thin spherical shell of radius $10^{16}$~cm, thickness $10^{14}$~cm,
and $H$-rich composition of solar metallicity
\citep[$\log(\mathrm{Na/H}) + 12 = 6.17$;][]{asplund05} undergoing complete
recombination of \ion{Na}{1}, similar to the treatment of SN~2006X in
\citet{patat07}, we obtain a CSM shell mass $\Menv < 3 \times 10^{-7}$~\Msol.
Similar limits can be obtained by multiplying the upper limit on the hydrogen
column density by the surface area of a sphere of radius $10^{16}$~cm
($6 \times 10^{-7}$~\Msol), or by using the estimated fluence of ionizing
photons from \citet{patat07} ($6 \times 10^{-7}$~\Msol).  The known H-rich
SNe~Ia-CSM, such as SN~2005gj, have estimated electron densities and CSM
masses several orders of magnitude higher \citep{snf2005gj}, as do the total
CSM masses ejected in the tidal tails of the mergers simulated by
\citet{rk13}.

These estimates, of course, assume hydrogen-rich CSM, whereas shock-powered
models for super-Chandrasekhar-mass SNe~Ia posit CSM rich in helium or even
carbon \citep{hachinger12,taub13}.  The first ionization potentials for
carbon (11.3~eV) and oxygen (13.6 eV) are comparable to that of hydrogen,
so one might expect similar electron densities from photoionization in those
cases; however, the expected relative abundance of sodium in such material
is highly uncertain, making it difficult to set definite limits.  For helium
the ionization potential is much higher (24.6~eV), requiring hard UV flux
blueward of 500~\AA; this entirely precludes useful limits from \ion{Na}{1}~D
absorption for CSM composed predominantly of helium.

To summarize, we have compiled the following lines of evidence regarding
CSM interaction in LSQ12gdj:
\begin{enumerate}
\item Since LSQ12gdj is clearly typed as a SN~Ia near maximum light,
      any CSM by this time either must be optically thin or must not cover
      the entire photosphere.  The fraction of luminosity which can be
      produced by shock heating or other non-radioactive sources is
      limited to about 75\% of the total \citep{leloudas13}.
\item The weak \ion{Na}{1}~D absorption limits the mass of extended
      hydrogen-rich CSM around LSQ12gdj to be less than $\sim 10^{-6}$~\Msol.
\item An extended all-helium or carbon-oxygen CSM could in theory evade
      the \ion{Na}{1}~D constraints, but would probably produce a lingering
      power-law tail to the light curve, as in SN~2005gj or PTF11kx,
      which we do not see in LSQ12gdj.
\item Fits to the bolometric light curve of LSQ12gdj limit the size of any
      compact envelope to be $< 10^{13}$~cm.  In this case the interaction
      would be frozen out before the first detection, resulting in all
      intermediate-mass elements being swept up into a reverse-shock shell
      and producing the very low velocity gradient observed.
\item If LSQ12gdj has a compact envelope, its relatively high \ion{Si}{2}
      velocity implies a light envelope of mass $\sim 0.1$~\Msol;
      this traps some radiation, but not as much as might be trapped in
      a heavily enshrouded explosion.
\end{enumerate}
It seems therefore that while some CSM may be present around LSQ12gdj,
luminosity from ongoing shock interaction is negligible.
Without tell-tale emission lines, however, the composition of the CSM and
the evidence for a single-degenerate origin for LSQ12gdj remain ambiguous.


\subsection{Implications for SN~2007if and SN~2009dc}
\label{subsec:2007if}

LSQ12gdj was flagged early in its evolution as a bright, peculiar SN~Ia.
By considering the UV contribution to LSQ12gdj's luminosity, we have shown
that up to 10\% of LSQ12gdj's maximum-light luminosity may be trapped thermal
energy from an interaction with a compact envelope.  Such a model, with small
variations in the relative contributions of \nickel\ mass and radioactivity
to the maximum-light luminosity, can explain the observational appearance
of 1991T-like SNe~Ia of comparable luminosity,
including SN~1991T itself and the SNe~Ia analyzed in \citet{scalzo12}.

We now consider what lessons may extend to the much brighter SNe~Ia,
SN~2007if and SN~2009dc, if any.  \revised{
While \citet{taub11}, \citet{hachinger12}, and \citet{taub13} considered a
number of possible physical scenarios for SN~2009dc, they were led to present
a white dwarf exploding inside an envelope as the most likely scenario based
on the following considerations:  The abundance patterns derived from
photospheric-phase and nebular-phase spectra are characteristic of the
thermonuclear explosion of a white dwarf, rather than of a core-collapse
event.  A single rapidly rotating white dwarf with the necessary mass ($> 2
\Msol$) and inferred \nickel\ mass would have been difficult to explain from
the standpoint of binary star evolution.  Violent mergers or collisions in
double-degenerate systems are expected to produce highly asymmetric
explosions, while the lack of continuum polarization implies a spherically
symmetric event \citep{tanaka10}.  \revised{Requiring that the event be
spherical also rules out models which explain SN~2009dc's luminosity mainly
through viewing angle effects \citep{hsr07}.}
Finally, even those channels able to
produce very large \nickel\ masses, such as white dwarf collisions
\citep{raskin10,kushnir13}, produce ejecta velocities which are too high to
match the observations.  An interaction with an envelope converts kinetic
energy into luminosity, enabling a powerful explosion to have low ejecta
velocity \citep{scalzo10} and potentially relaxing the requirement of a very
high \nickel\ mass.}

\revised{However, \citet{hachinger12} and \citet{taub13} make no specific
predictions for the geometry or physics of the interaction.
\citet{hachinger12} show that spectra of SN~2009dc can be reproduced by the
sum of a SN Ia spectrum and a smooth pseudocontinuum; they consider
polynomials and spectra of SNe with strong CSM interactions as possible
functional forms.  The interaction luminosity is simply assumed to be the
difference between what is observed and what is predicted from radioactive
decay.  \citet{taub13} examine the influence of the post-interaction ejecta
density profile on radiation trapping at late times, and estimate a CSM mass
of about 0.6~\Msol; they make few predictions about the CSM geometry necessary
to reproduce the near-maximum light curve, and raise concerns about
fine-tuning.}

\revised{Using LSQ12gdj as a point of departure, we can reason
about how the presence of interaction luminosity affects inferences about
the ejected masses and \nickel\ masses of SN~2007if and SN~2009dc.
While ongoing interactions may run into fine-tuning problems, an interaction
can contribute to maximum-light luminosity while leaving the late-time light
curve undisturbed as long as the envelope is sufficiently compact.}

\revised{The influence of interaction with a compact envelope can thus be
crudely estimated as an adjustment to the luminosity-to-radioactivity ratio
$\alpha$.  The \citet{scalzo10} analysis of SN~2007if}
used $\alpha = 1.3 \pm 0.1$, i.e., it assumed that around 30\% of SN~2007if's
maximum-light luminosity was trapped radiation, and a long rise time of
23~days.  While the rising part of the light curve was well-sampled by
ROTSE-III \citep{yuan10}, with the first detection at 20~days before
$B$-band maximum light, SN~2007if has only one pre-maximum bolometric
light curve point, making its maximum-light colour uncertain \citep{scalzo12}
and precluding a more detailed analysis of the pre-maximum light curve.
Crucially, SN~2007if also has no UV data.  If the UV component of SN~2007if's
bolometric luminosity evolved in a similar way to LSQ12gdj's, this would have
made SN~2007if 17\% more luminous at peak ($3.7 \times 10^{43}$~erg~s$^{-1}$),
requiring a \nickel\ mass of around ($2.0/\alpha$)~\Msol.
Most of the trapped radiation should be gone around 60~days after explosion,
so the late-time light curve measurements correctly reflect that the ejecta
must have been extremely massive.
If we assume $\alpha = 2.0$, bringing \MNi\ down to 1.0~\Msol, the limit
of what can be achieved in a Chandrasekhar-mass explosion \citep{kmh93},
we must still have $\MWD > 2.26$~\Msol\ at 99\% confidence.

Similar considerations apply to SN~2009dc, which has an almost identical
light curve to SN~2007if out to 100~days past maximum light.  SN~2009dc also
has \emph{Swift} data \citep{silverman11}, though none before maximum light,
so the precise shape of its pre-maximum bolometric light curve is still
subject to large uncertainties.  At maximum light, \citet{silverman11}
estimate that about 20\% of SN~2009dc's bolometric flux is emitted in the UV,
similar to LSQ12gdj.  The low absorption-line velocities make it impossible
for SN~2009dc to have a ``normal'' density structure, or even much 
burned material beyond about 9000~\kms.  For SN~2009dc to have been a tamped
detonation \revised{or pulsating delayed detonation, the outer, incompletely
burned layers of ejecta} must have represented a much larger
fraction of the total ejecta mass --- possibly as high as 30\% of the total
--- in order to reproduce its even lower ejecta velocities.
Under these conditions, the reverse shock should penetrate far into the
inner layers of ejecta before stalling, and the distribution of material
in the reverse-shock shell becomes important to gamma-ray transport at late
times.  Thus the approximation previously used by \citet{scalzo12} for
SN~2007if and other super-Chandrasekhar-mass candidates, in which the shock
redistributes kinetic energy and traps thermal energy but has little effect
on the late-time light curve, probably breaks down for SN~2009dc.
\revised{Detailed hydrodynamic simulations of explosions inside compact
envelopes could be used to suggest a suitable density profile.}

\revised{In contrast to SN~2007if,}
SN~2009dc's relatively large ($\sim 75$~\kms~day$^{-1}$) \ion{Si}{2} velocity
gradient presents a problem for \revised{explosion scenarios which produce
shell structures in the ejecta \citep{qhw07},} because no \revised{velocity}
plateau is evident.  It is also clear that intermediate-mass elements cannot
all be trapped in a thin layer, \revised{as with the delayed detonation
scenario suggested by \citet{childress13} for SN~2012fr.}
However, given the strength of the shock necessary, the approximation of the
reverse-shock shell as a thin layer could also break down here.
A pulsating delayed detonation could have given SN~2009dc a highly
disturbed density structure without the need for an envelope or for a very
narrow layer of intermediate-mass elements in velocity space.  \citet{baron12}
invoked such a model for the slow-declining SN~2001ay \citep{krisciunas11}.

One difficulty with pulsating models for SN~2007if and SN~2009dc is that,
\revised{while the pulsation will thermalize and redistribute kinetic energy,}
the shock freezes out much sooner after explosion than in the case of
a tamped detonation.  This forces $w = 0$ and prevents a significant
contribution \revised{of trapped thermalized kinetic energy} to the
maximum-light luminosity (but potentially further enhancing trapping of
radioactive energy near maximum light).

\section{Conclusions}

LSQ12gdj is a well-observed, overluminous SN~Ia in a nearby galaxy with
little to no dust extinction.  The extensive spectroscopic time series
show that LSQ12gdj is spectroscopically 1991T-like, with intermediate-mass
element absorption signatures only in a narrow range of velocities,
much like SN~2007if and other 1991T-like SNe~Ia \citet{scalzo10,scalzo12}.
From the bolometric light curve of LSQ12gdj we infer a \nickel\ mass of about
1.0~\Msol\ and an ejected mass near the Chandrasekhar mass.

Observations at UV wavelengths well before maximum light provide additional
useful constraints on the properties of LSQ12gdj and other 1991T-like SNe~Ia,
not considered elsewhere.  A large fraction (17\%) of the bolometric
luminosity near maximum light, and nearly 30\% in the earliest observations,
is emitted bluewards of 3300~\AA.  Accounting for this effect increases the
derived \nickel\ mass significantly relative to cases in which it is ignored
\citep[e.g.,][]{scalzo12}, assuming that the SN is powered through
radioactivity alone.

\revised{Our excellent time and wavelength coverage also allow us to consider
alternative sources of luminosity for LSQ12gdj, which can guide our intuition
for other luminous super-Chandrasekhar-mass SN~Ia candidates.  We find that
as much as 10\% of LSQ12gdj's luminosity could come from trapped thermal
energy from an early-phase shock interaction, with virtually none coming from
ongoing shocks at later times.  Such a mechanism could in principle
explain the extreme luminosities and low photospheric velocities of SN~2007if
and SN~2009dc as resulting from the trapping of thermalized kinetic energy
from a short interaction at early times, without appealing to ongoing shock
interactions with extended winds which are likely to cause greater deviations
from SN~Ia behavior than observed.}

Our findings represent what can be done with detailed observations, and to
push our understanding of super-Chandrasekhar-mass SNe~Ia forward,
even more detailed observations will be needed.  Early ultraviolet coverage
is critical, starting as soon after explosion as possible.  Optical and
near-infrared observations extending to late times, well past maximum light,
are needed to place helpful constraints on the mass.  Nebular spectra
can elucidate the density structure of the innermost ejecta, with implications
for the importance of radiation trapping near maximum light.
These observations must go hand in hand with sophisticated, self-consistent
modelling which can deal with theoretical uncertainties and with systematic
errors in the observations.

Measurement of the properties of a general spectroscopically selected
sample of 1991T-like SNe~Ia could provide vital clues to the identity of
their progenitors and how they relate to other super-Chandrasekhar-mass
SNe~Ia, such as SN~2006gz and SN~2009dc, and to CSM-interacting SNe~Ia,
such as SN~2005gj and PTF11kx.
\revised{\citet{leloudas13} show a strong association
between 1991T-like SNe~Ia and the growing Ia-CSM subclass which show narrow
H$\alpha$ lines in their spectra \citep{silverman13b}; they imply that
1991T-like SNe~Ia must in general be single-degenerate explosions, although
not all of them are required to display strong CSM interaction.}
\revised{Similarly,} a SN~Ia sample from an untargeted search selected only
by peak absolute magnitude can determine the spectroscopic diversity and
range of explosion mechanisms which can account for superluminous SNe~Ia,
and how many superluminous SNe~Ia result from
the explosions of super-Chandrasekhar-mass white dwarfs.


\section*{Acknowledgments}

PyRAF and PyFITS are products of the Space Telescope Science Institute,
which is operated by AURA for NASA.
This research has made use of the NASA/IPAC Extragalactic Database (NED)
which is operated by the Jet Propulsion Laboratory, California Institute of
Technology, under contract with the National Aeronautics and Space
Administration.
This research is based on observations collected at the European
Organisation for Astronomical Research in the Southern Hemisphere, Chile
as part of PESSTO (the Public ESO Spectroscopic Survey for Transient Objects),
ESO program ID 188.D-3003.
Research leading to these results has received funding from the European
Research Council under the European Union's Seventh Framework Programme
(FP7/2007-2013)/ERC Grant agreement n$^{\rm o}$ [291222] (PI : S. J. Smartt).
The National Energy Research Scientific Computing Center, supported by the
Office of Science of the U.S. Department of Energy under Contract No.
DE-AC02-05CH11231, provided staff, computational resources, and data storage
for this project.
Parts of this research were conducted by the Australian Research Council
Centre of Excellence for All-Sky Astrophysics (CAASTRO), through project
number CE110001020.
This material is also based upon work supported by NSF under
grants AST--0306969, AST--0607438 and AST--1008343.
RS acknowledges support from ARC Laureate Grant FL0992131.
ST acknowledges support from the Transregional Collaborative Research Center
TRR 33 ``The Dark Universe'' of the Deutsche Forschungs\-gemeinschaft.
KM is supported by a Marie Curie Intra European Fellowship, within the 7th
European Community Framework Programme (FP7).
MF is supported by the European Union FP7 programme through ERC grant number
320360.
MS and CC gratefully acknowledge generous support provided  by the Danish
Agency for Science and Technology and Innovation
realized through a Sapere Aude Level 2 grant. 
AG acknowledges support by the EU/FP7 via ERC grant no. 307260, a GIF grant,
the Minerva ARCHES award and the Kimmel award.


\end{document}